%% file: mlvector.tex
\documentclass[preprint,floatfix,prb,superscriptaddress,nofootinbib]{revtex4-1}
\usepackage[T1]{fontenc}
\usepackage[utf8]{inputenc}
\usepackage{graphicx}
\usepackage{overpic}
\usepackage{amsmath}
\usepackage{color}
\usepackage{comment}
\usepackage{hyperref}
\usepackage{multirow}
\usepackage{subfigure}
\usepackage{etoolbox}

\input{short.tex}

\makeatletter
\def\@email#1#2{%
 \endgroup
 \patchcmd{\titleblock@produce}
  {\frontmatter@RRAPformat}
  {\frontmatter@RRAPformat{\produce@RRAP{*#1\href{mailto:#2}{#2}}}\frontmatter@RRAPformat}
  {}{}
}%
\makeatother

\begin{document}

\title{A simple approach to rotationally invariant machine learning of a vector quantity}

\author{Jakub Martinka}
\affiliation{J. Heyrovsk\'{y} Institute of Physical Chemistry, Academy of Sciences of the Czech \mbox{Republic, v.v.i.}, Dolej\v{s}kova 3, 18223 Prague 8, Czech Republic}
\affiliation{Department of Physical and Macromolecular Chemistry, Faculty of Sciences,
Charles University, Hlavova 8, 128 43 Prague 2, Czech Republic}

\author{Marek Pederzoli}
\affiliation{J. Heyrovsk\'{y} Institute of Physical Chemistry, Academy of Sciences of the Czech \mbox{Republic, v.v.i.}, Dolej\v{s}kova 3, 18223 Prague 8, Czech Republic}

\author{Mario Barbatti}
\affiliation{Aix Marseille University, CNRS, ICR, Marseille, France}
\affiliation{Institut Universitaire de France, 75231 Paris, France}

\author{Pavlo O. Dral}
\affiliation{State Key Laboratory of Physical Chemistry of Solid Surfaces, College of Chemistry and Chemical Engineering, and Fujian Provincial Key Laboratory of Theoretical and Computational Chemistry, Xiamen University, Xiamen, Fujian 361005, China}
\affiliation{Institute of Physics, Faculty of Physics, Astronomy, and Informatics, Nicolaus Copernicus University in Toruń, ul. Grudziądzka 5, 87-100 Toruń, Poland}

\author{Ji\v{r}\'{i} Pittner}%
\email{jiri.pittner@jh-inst.cas.cz}
\affiliation{J. Heyrovsk\'{y} Institute of Physical Chemistry, Academy of Sciences of the Czech \mbox{Republic, v.v.i.}, Dolej\v{s}kova 3, 18223 Prague 8, Czech Republic}

\begin{abstract}
Unlike with the energy, which is a scalar property, machine learning (ML) predictions of vector or tensor properties poses the additional challenge of achieving proper invariance (covariance) with respect to molecular rotation. For the energy gradients needed in molecular dynamics (MD), this symmetry is automatically fulfilled when taking analytic derivative of the energy, which is a scalar invariant (using properly invariant molecular descriptors). However, if the properties cannot be obtained by differentiation, other appropriate methods should be applied to retain the covariance.
There have been several approaches suggested to properly treat this issue. For nonadiabatic couplings and polarizabilities, for example, it was possible to construct 
virtual quantities from which
the above tensorial properties are obtained by differentiation and thus guarantee the covariance. Another possible solution is to build the rotational equivariance into the design of a neural network employed in the model. 
Here we propose a simpler alternative technique, which does not require construction of auxiliary properties or application of special equivariant ML techniques. We suggest a three-step approach, using the molecular tensor of inertia. In the first step, the molecule is rotated using the eigenvectors of this tensor to its principal axes.
In the second step, the ML procedure predicts the vector property relative to this orientation, based on a training set where all vector properties were in this same coordinate system. 
As third step, it remains to transform the ML estimate of the vector property back to the original orientation. This rotate-predict-rotate (RPR) procedure should thus guarantee proper covariance of a vector property and is trivially extensible also to tensors such as polarizability.
The PRP procedure has an advantage that the accurate models can be trained very fast for thousands of molecular configurations which might be beneficial where many trainings are required (e.g., in active learning).
We have implemented the RPR technique, using the MLatom and Newton-X programs for ML and MD and performed its assessment on the dipole moment along MD trajectories of \mbox{1,2-dichloroethane}.
\end{abstract}

\maketitle

\section{Introduction}
Machine learning (ML) proved to be a powerful tool in computational chemistry\citep{Dral2020}. While ML prediction of scalar quantities, such as potential energy to propagate molecular dynamics (MD) simulations, is a \mbox{well-established} procedure\citep{Lorenz2004,Raff2005}, predicting vectors (rank-1 tensors) or higher rank tensors challenge the ML techniques to deliver quantities covariant with respect to symmetry operations\citep{Dral_book_19}.
This problem affects various properties such as dipole moment, (hyper)polarizability or forces, which are crucial for spectrum (infrared\citep{Perakis2016}; Raman\citep{Zhang2016}), machine learning potentials (MLPs)\citep{Cheng2024} and MD simulations. Other vectorial properties possess yet an additional demand of having a correct sign. This is the case of transition dipole moments\citep{Ye2019,Zhang2021,Ye2020,Westermayr2019,Westermayr2020a} and nonadiabatic coupling vectors\citep{Westermayr2019,Westermayr2020b,Richardson2023}, which are calculated for two different electronic states of an arbitrary global phase.

There are several approaches in the literature to handle orientation dependence. For dipole moments, neural networks were suggested that learn atomic charges to reconstruct the dipole moments as accurately as possible\citep{PhysNet2019,VSpecNN2024}. For polarizabilities and nonadiabatic couplings, such neural networks were built that can learn the virtual quantities, the analytical derivatives of which produce the required tensorial properties\cite{Westermayr2020b,Zhang2020,VSpecNN2024}. 
Similar in spirit is an approach based on kernel methods which was used to learn dipole moments as response properties.~\cite{christensen2019} Above approaches are inspired by MLPs where it is common to derive energy gradients (negative forces) as the derivatives of the machine learning function of potential energies.
Many approaches rely on building equivariance into the design of the ML model, either into the local descriptors employed in neural network (NN) approaches, or in the design of the NN itself. This was done in case of Gaussian process regression (GPR), by starting with a kernel  based on a smooth overlap of atomic positions (SOAP)\citep{Bartok2013} and deriving a hierarchy of $\lambda$-SOAP kernels preserving rotational symmetry\citep{Glielmo2017,Grisafi2018,Wilkins2019}. This so-called symmetry-adapted GPR (SA-GPR) approach was used to paracetamol system leading to a covariant predictions\citep{Raimbault2019}.
The same is possible for the class of ML algorithms based on NNs as shown for neuroevolution potential (NEP) scheme\citep{Fan2021}, which was modified into tensorial-NEP approach\citep{Xu2024} or embedded atom NN (EANN)\citep{Zhang2019} transformed into tensorial EANN,\citep{Zhang2020} outperforming SA-GPR.
Other examples introduced new rotationally covariant architectures\citep{Anderson2019}, equivariant message passing NN\citep{Schutt2021} or tensor field NN\citep{Thomas2018}. The NN-based approaches proved to be excellent tools for predicting infrared and Raman spectra\citep{Chen2024,Zhang2020a}, even under the influence of external fields\citep{Zhang2023a}.

While above approaches all shown to work well for their test applications, the practical problem is their often high computational cost for training and predicting. The solution to this problem is often employing the kernel methods and, indeed, the kernel ridge regression (KRR) with a global molecular descriptor has already been suggested for treatment of such tensorial properties as multipole moments \cite{bereau2015b} and atomic forces \cite{rupp2015}.
In such approaches, it is common to rotate the molecular geometry to some reference frame\citep{Bereau2015,Liang2017,HuHuo2023}.
Although such a rotation is straightforward to implement, care should be taken for flexible or dissociative systems\citep{Grisafi2018}. 
Here we built upon this rotational idea to propose a three-step procedure to maintain proper covariance in the context of learning the vectorial properties for the set of conformations visited during molecular dynamics. 
Our procedure is similar in spirit to the one employed by Hu and Huo \cite{HuHuo2023} in a related context, however, we define the reference frame 
by principal axes of the molecular tensor of inertia rather than by an arbitrarily selected triple of atoms, which should remove a potential source of ambiguity.
We demonstrate this new method (named rotate-predict-rotate for reason that will be clear in the next section) on 1,2-dichloroethane, for which new descriptors were introduced improving the overall test set root mean square error (RMSE) values for both dipole moment and polarizability components.
The proposed method is not just accurate but also computationally very efficient, which we demonstrate by timings for different sizes of the learning data set.

\section{Methods}
In this study, we used kernel ridge regression (KRR) to fit individual components of vectorial/tensorial properties, which are predicted using 
\begin{equation}
    y(\bx') = \sum_{i=1}^{N_\text{tr}}\alpha_iK(\bx',\bx_i),
\end{equation}
where $N_{\text{tr}}$ is a number of training points, $\alpha_i$ is a regression coefficient and, $K$ represents kernel function which can has Gaussian (\ref{eq:Gauss}), Laplace (\ref{eq:Laplace}) or Mat\'{e}rn (\ref{eq:Matern}) form
\begin{subequations}
    \begin{equation}\label{eq:Gauss}
        K({\bx}_i, {\bx}_j)=\exp\left(\frac{-||{\bx}_i-{\bx}_j||^2}{2\sigma^2}\right),
    \end{equation}
    \begin{equation}\label{eq:Laplace}
        K({\bx}_i, {\bx}_j)=\exp\left(\frac{-||{\bx}_i-{\bx}_j||}{\sigma}\right),
    \end{equation}
    \begin{equation}\label{eq:Matern}
        K({\bx}_i, {\bx}_j)=\sum_{k=0}^n\frac{(n+k)!}{(2n)!} \left( \frac{2||{\bx}_i-{\bx}_j||}{\sigma}\right)^{n-k}\exp\left(\frac{-||{\bx}_i-{\bx}_j||}{\sigma}\right).
    \end{equation}
\end{subequations}

The geometrical configuration of a molecule is represented by a molecular descriptor $\bx$. There is a plethora of descriptors, usually suitable for various tasks. 
Common examples of global descriptors suitable for use with KRR are relative to equilibrium (RE, (\ref{eq:RE})), Coulomb matrix (CM, (\ref{eq:CM})) or inverse distance (ID, (\ref{eq:ID}))
\begin{subequations}
    \begin{equation}\label{eq:RE}
        {\bx}^\top_{\text{RE}}=\left( \frac{r^{\rm eq}_{12}}{r_{12}},\cdots, \frac{r^{\rm eq}_{23}}{r_{23}},\cdots, \frac{r^{\rm eq}_{(N_{\rm at}-1)N_{\rm at}}}{r_{(N_{\rm at}-1)N_{\rm at}}} \right),
    \end{equation}
    \begin{equation} \label{eq:CM}
        {\bx}^\top_{\text{CM}}=\left( 0.5Z_1^{2,4},\frac{Z_1Z_2}{r_{12}},\cdots, \frac{Z_2Z_1}{r_{21}}, 0.5Z_2^{2,4},\cdots\right),
    \end{equation}
    \begin{equation} \label{eq:ID}
        {\bx}^\top_{\text{ID}}=\left(\frac{1}{r_{1,2}},\frac{1}{r_{1,3}},\cdots\right).
    \end{equation}
\end{subequations}

In the training, the regression coefficients are obtained by analytically solving the system of equations in matrix form $(\bK+\lambda I)\balpha=\by$. For each trainig, we also optimize the hyperparameters $\lambda$ and $\sigma$ (and in case of Mat\'{e}rn kernel also $n$) by using the 20\% of the training set as the validation set). 
{The optimization of the hyperparameters was performed for each vector (or tensor) component independently.}
While there is a wide range of possible combinations of descriptors and kernels within KRR, \verb|RE| descriptor with Gaussian kernel (the KREG model\citep{Dral2017,Hou2023}) proved to be very accurate in learning potential energy surfaces (PESs) \citep{Pinheiro2021,Hou2023}.

If individual components of a vector or tensor quantity were fitted separately in this way,
a proper covariance (transformation with respect to molecular rotations) would not in general be guaranteed.
As a simple approach to achieve rotationally covariant results in the KRR method, 
we suggest the rotate-predict-rotate (RPR), a three\mbox{-}step approach based on the molecular tensor of inertia
\begin{equation}
    I_{ij} = \sum^{\text{atoms}}_\alpha m_\alpha(||\br^\alpha ||^2 \delta_{ij} - x^\alpha_i x^\alpha_j),
\end{equation}
where $\br^\alpha=\left[ x_1^\alpha,x_2^\alpha,x_3^\alpha \right]$ 
is the radius\mbox{-}vector of atom $\alpha$ with respect to the center of mass (alternatively, atomic charges and the center of charge can be employed too). 
This is a symmetric positive definite matrix and its diagonalization yields transformation matrices $\bQ$ from current orientation to the principal axes of inertia
\begin{equation}
    {\bI} = {\bQ\boldsymbol{\Lambda}\bQ}^\top,
\end{equation}
where $\bQ$ is the orthogonal rotation matrix and $\bf \Lambda$ is a diagonal matrix of principal moments of inertia. 

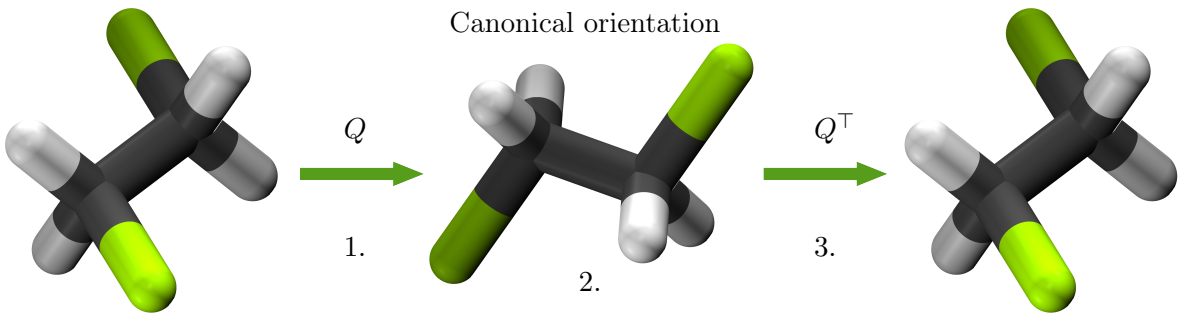
\begin{figure}
\centering
\begin{overpic}[width=.95\textwidth]{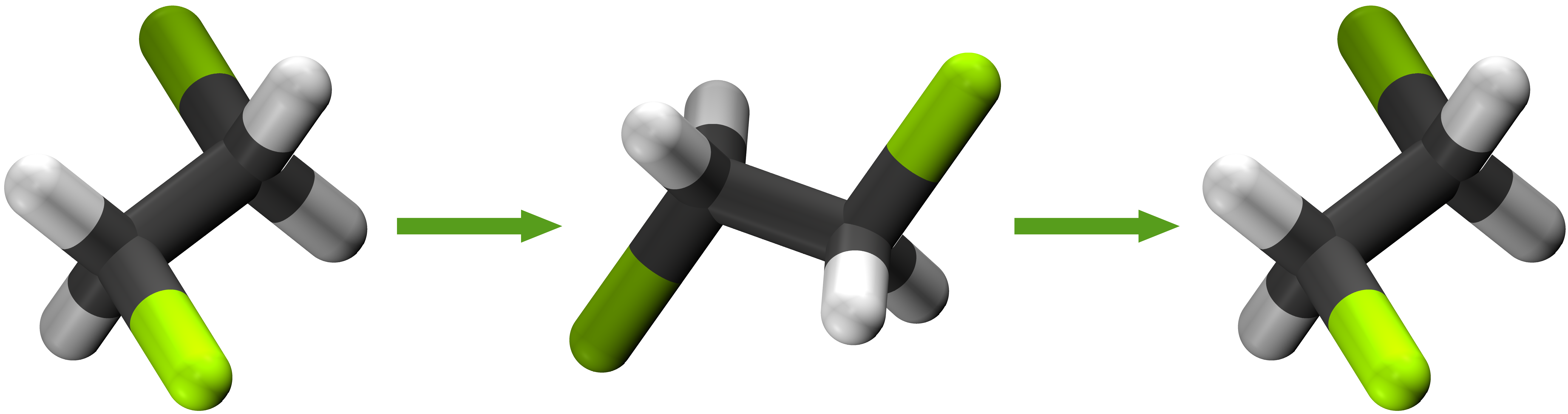}
    \put(29,15){$Q$}
    \put(29,5){$1.$}
    \put(69,15){$Q^\top$}
    \put(69,5){$3.$}
    \put(49,2){$2.$}
    \put(38,24){Canonical orientation}
\end{overpic}
    \caption{Proposed RPR approach: 1. rotation of molecule and corresponding properties by $\bQ$, 2. prediction by a model trained on molecules rotated into their canonical orientation, 3. backward rotation into the initial orientation $\bQ^\top$.}
    \label{fig:rotation}
\end{figure}

In the first step, the molecule is rotated using $\bQ$ to its principal axes. I.e., if some vector property was already computed in the original orientation, it is rotated to the new canonical orientation before being used for ML training. 
In the second step, the ML procedure predicts the vector property relative to the canonical orientation, based on a training set where all vector properties were in this same coordinate system. 
In the third step, the ML estimate of the vector property is transformed back to the original orientation using $\bQ^{-1} = \bQ^\top$. The RPR procedure, depicted in Fig.~\ref{fig:rotation}, should thus guarantee proper covariance of a vector property.
Notice that thanks to the uniquely defined canonical orientation, the molecular descriptor itself does not need to be invariant to rotations.
If fact, we tried even the set of cartesian coordinates with atomic charges as a descriptor and the covariance was properly maintained by the RPR procedure.

For higher rank tensors, the RPR procedure is easily generalized by application of the $\bQ$-transformation
to each of its indices.
When the general linear transformations of tensors are restricted to proper rotations, the cartesian tensors further split according to the irreducible representations of the SO(3) group. 
In particular, the second rank symmetric polarizability tensor decomposes to scalar polarizability being 1/3 of the trace and a traceless spherical tensor of rank 2 with 5 independent components. 
In the RPR procedure, we performed transformations of the original cartesian tensor, but verified by means of scatter plots also the results of the ML procedure on the aforementioned spherical components. 

\section{Numerical results}
To numerically verify the RPR procedure, we have chosen \mbox{1,2-dichloroethane}, which has
a dipole moment and polarizability strongly dependent on the \mbox{Cl-C-C-Cl} dihedral angle. 
This makes it an ideal test system for ML of these properties.
To assemble the training and testing data sets, we performed MD of this molecule in its ground state.
For this purpose, the \mbox{Newton-X}\citep{NewtonX2022,NewtonX2013} program, interfaced with Turbomole 7.3\citep{TURBOMOLE_7.3}, was used to compute the MD trajectories. 
All single point calculations were carried out using density functional theory (DFT) with the B3LYP functional and \mbox{aug-cc-pVTZ} basis set.
The initial conditions were obtained from a harmonic-approximation Wigner distribution with temperature of 298 K, which was maintained during the MD using the Andersen thermostat. Geometries were then modified by varying the dihedral angle between \mbox{Cl-C-C-Cl} atoms by 5 degrees to yield geometries covering the whole dihedral range {from 0 to 180 degrees}.
{The change of the dihedral angle was performed by converting the Wigner's initial conditions to internal coordinates and changing the dihedral angle by an appropriate increment of a multiple of 5 degrees, while the other internal coordinates and the velocities were unchanged. While this procedure is somewhat artificial, its purpose was not to produce a physically motivated statistical ensamble, but to generate data for ML with widely varying \mbox{Cl-C-C-Cl} dihedral angles and consequently dipole moments.}

In total we propagated seventy trajectories with a duration of 50 fs each, using a time step of 0.5 fs. 
This resulted in a dataset of 7,000 points (consisting of geometries, energies, dipole moments, and polarizability tensors).
To analyze the learning behaviour, smaller training subsets were created by random selection from the full dataset.
Each component is fitted by an independent model trained on a dataset of geometries, which were rotated, together with the corresponding properties, using rotational matrices $\bQ$ to the canonical orientation.

Firstly, we started with a benchmark of various combinations of descriptors and kernels to compare their performance.
At this stage, all models were trained on a 1000-point dataset and evaluated on test set consisting of 1000 points selected from the trajectories.
We employed the descriptors already implemented in MLatom 2\citep{MLatom2019,MLatom2021}, i.e. \verb|RE|, \verb|CM|, \verb|ID|. 
However, in addition to these well-established descriptors, new descriptors tailored for this molecule were introduced by incorporating cartesian or spherical coordinates of chlorine atoms (\verb|RExyzCl| or \verb|REsphCl|), dihedral angles (\verb|REdihedCl|), or simply by utilizing cartesian coordinates multiplied by atomic charges (\verb|xyzQ|); cartesian coordinates were taken in the canonical orientation.
This was done to reflect the importance of the chlorine atoms since their orientation is crucial for the studied properties of \mbox{1,2-dichloroethane}. 
The addition of chlorine coordinates to the descriptor turned out to have a large influence, where predictions of $y$ and $z$ components of the dipole moment with the \verb|RExyzCl| descriptor produced RMSE values 20--40 times lower than the \verb|RE| descriptor.
This improvement was true for other descriptors as well, and the \verb|RExyzCl| descriptor also outperforms \verb|RE| in case of the polarizability components.
Concerning the kernel function, Mat\'{e}rn kernel performed the best for dipole moment components, while the Gaussian kernel was superior for polarizability components.

\begin{figure}
    \centering
    \includegraphics[width=.45\textwidth]{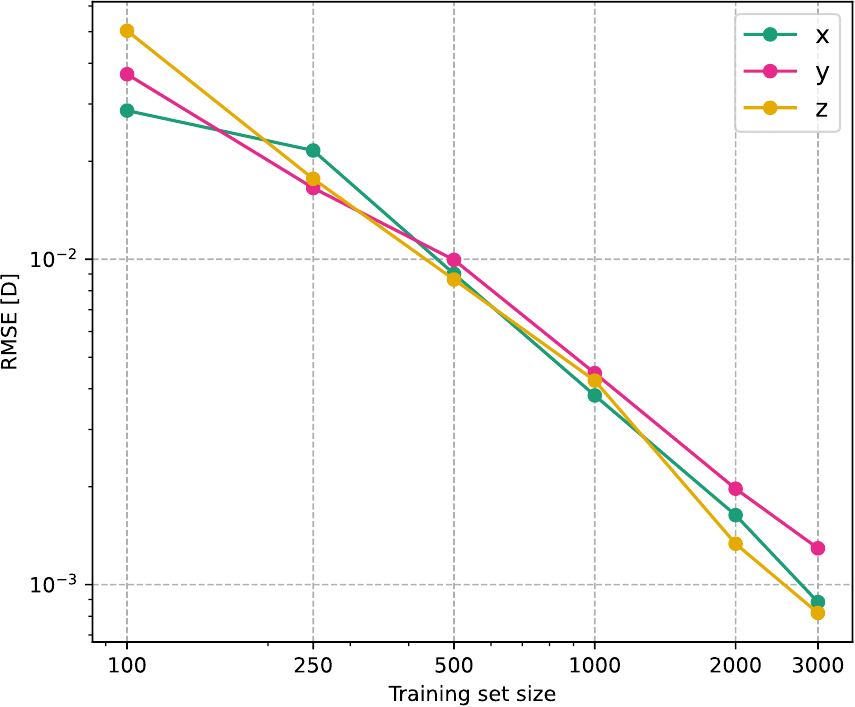}
    \includegraphics[width=.45\textwidth]{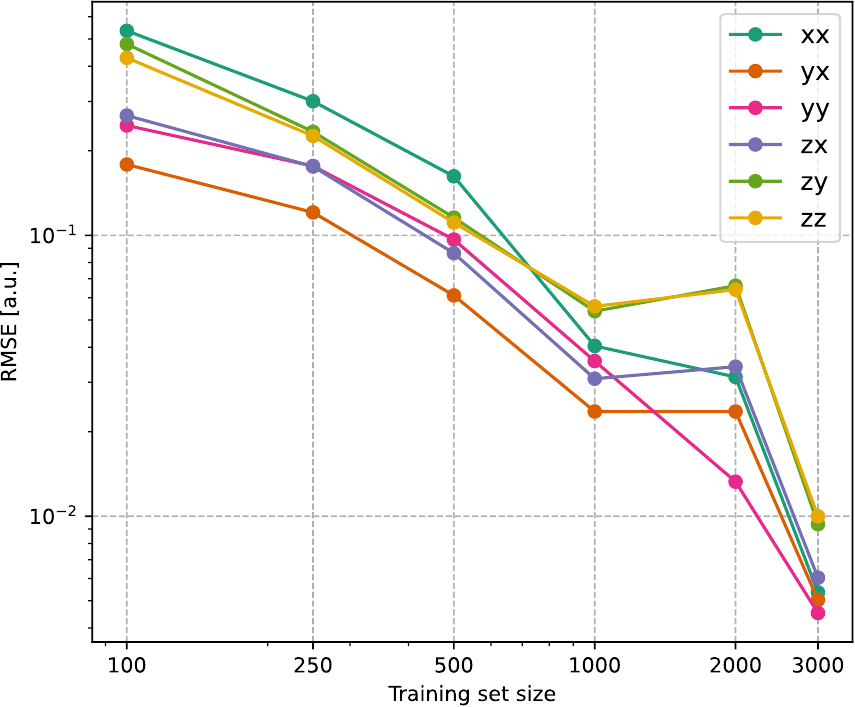}
    \caption{Learning curves of dipole moment components in Debye (left) and polarizability components in atomic units (right). Accuracy of each model was determined on the test set consisting of 1000 points.}
    \label{fig:learning_curves}
\end{figure}

The learning behaviour is shown in Fig.~\ref{fig:learning_curves} for the dipole moment as well as for the polarizability. 
It can be seen that the training set consisting of 1000 points already achieved satisfactory accuracy and will thus be discussed in the following sections.

\begin{figure}
    \centering
    \includegraphics[height=.4\textwidth]{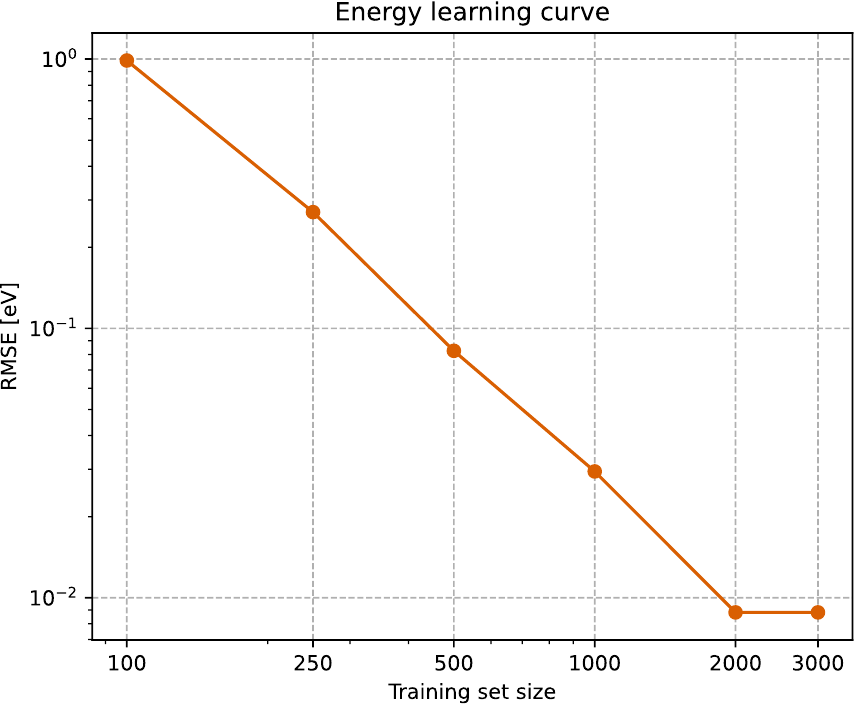}
    \includegraphics[height=.4\textwidth]{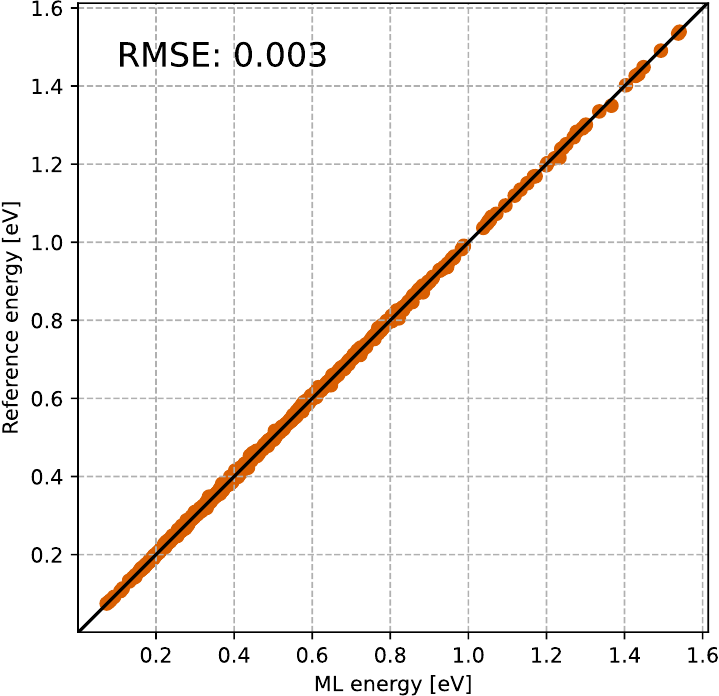}
    \caption{Learning curve and scatter plot of a KREG model of $S_0$ state trained on a dataset consisting of 1000 points. This model was used to propagate 24 trajectories in order to construct the test set.}
    \label{fig:kreg}
\end{figure}

To validate the proposed approach, an independent test set covering all spatial orientations as well as dihedral angle values is required, since points of a test set constructed from the original dataset are strongly correlated with the training data, thus leading to overoptimistic RMSE.
To achieve this, we fitted the ground state PES using the KREG model\citep{Dral2017,Hou2023} and used \mbox{Newton-X} interfaced with MLatom to run multiple trajectories with the appropriate initial conditions. 
The KREG models were trained on the energies and energy gradients of the 1000-point dataset randomly selected from the 7000 training data points, for which learning curve and scatter plot in Fig.~\ref{fig:kreg} were constructed using a test set of 1000 points.

\begin{figure}
    \centering
    \includegraphics[width=.9\textwidth]{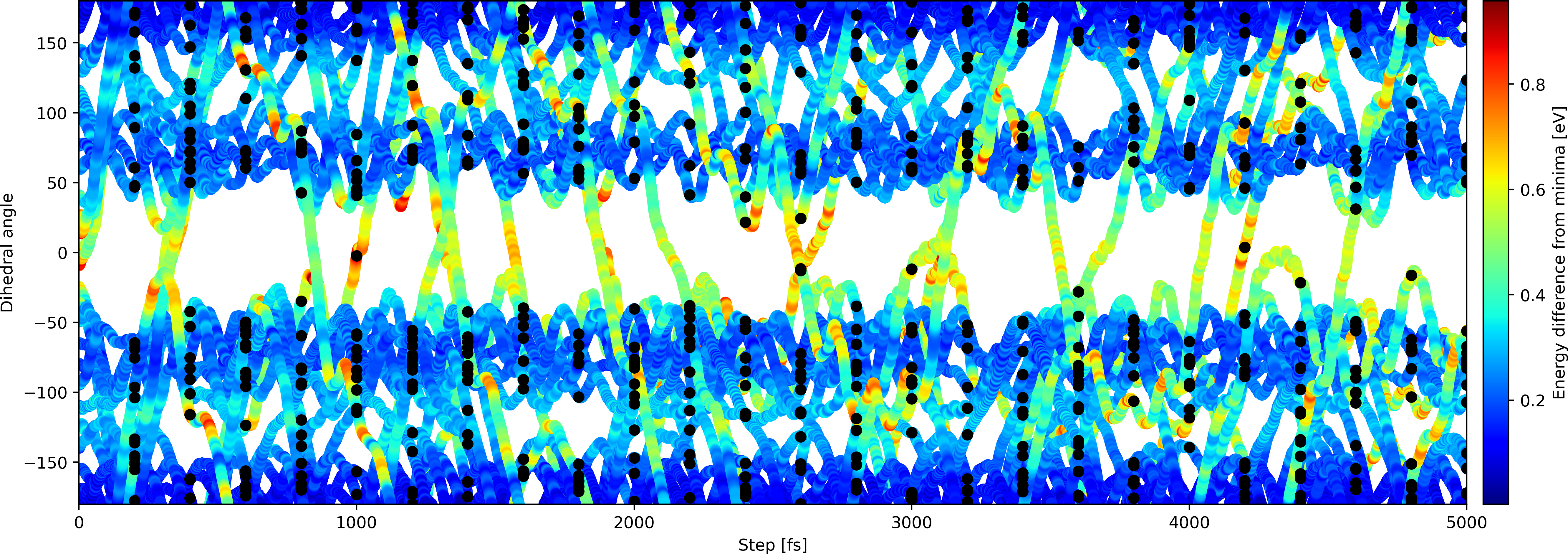}
    \caption{Dihedral angle of \mbox{Cl-C-C-Cl} with respect to simulation time. Black points were taken as a test set consisting of 1200 points, for which dipole moments/polarizability tensors were calculated using DFT and accuracy of trained models was determined.}
    \label{fig:dihedral}
\end{figure}

Sampling of the dihedral angle by 15 degrees led to 24 initial geometries. A Boltzmann distribution with a temperature of 300 K was used to generate velocities, and an additional uniformly random rotation of the molecule as a whole was imposed to properly sample all molecular orientations.
With a time step of 0.5 fs and a final time of 5 ps, a total of 240,000 points were obtained. 
This led to a dataset covering all molecular orientations in space and dihedral angles, which can be seen in Fig.~\ref{fig:dihedral}, showing the time evolution of the \mbox{Cl-C-C-Cl} dihedral angle.

\begin{figure}
    \subfigure[Dipole moment]{\includegraphics[height=.4\textwidth]{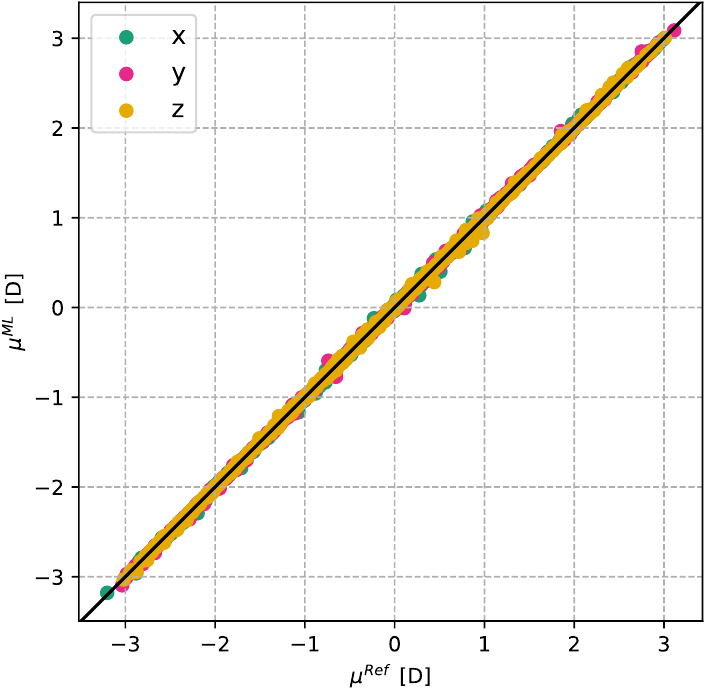}\label{fig:dip}}
    \subfigure[Polarizability]{\includegraphics[height=.4\textwidth]{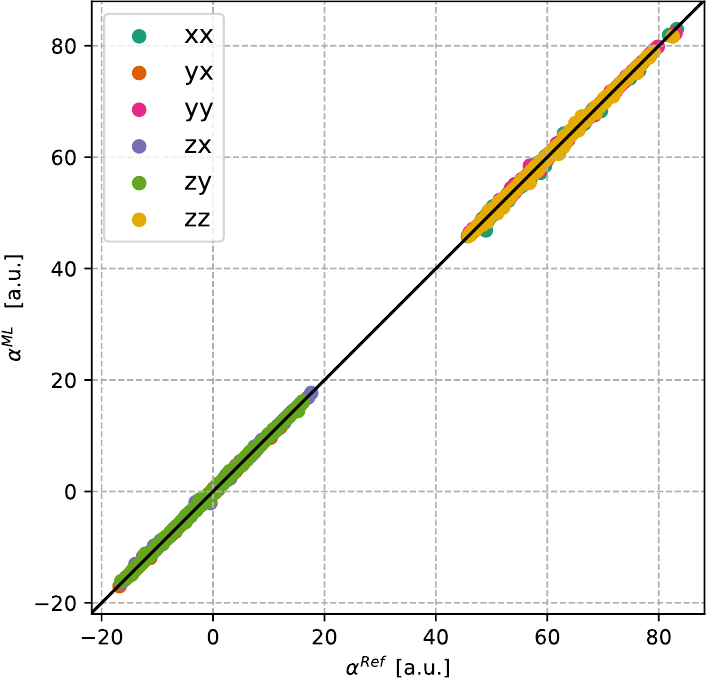}\label{fig:pol1}}\\
    \subfigure[Scalar polarizability]{\includegraphics[height=.4\textwidth]{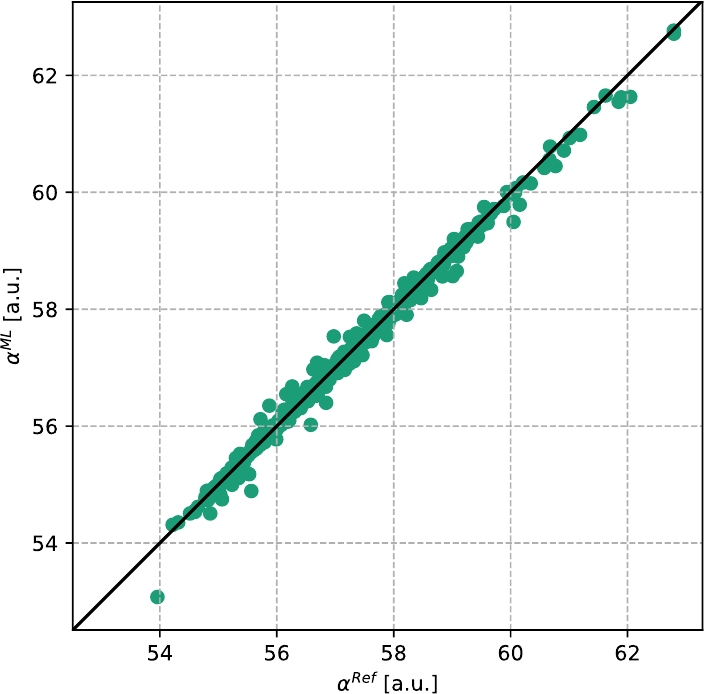}\label{fig:pol2}}
    \subfigure[Spherical polarizability]{\includegraphics[height=.4\textwidth]{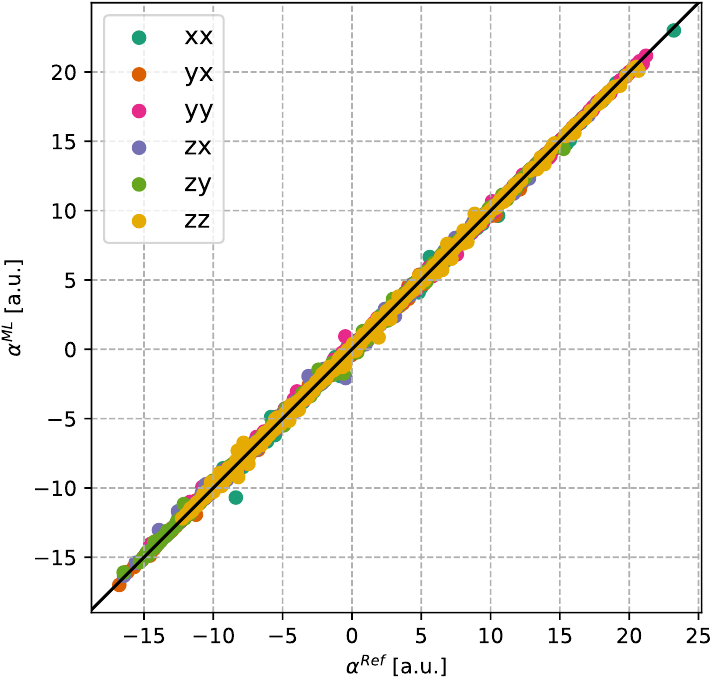}\label{fig:pol3}}
    \caption{Scatter plots showing the DFT dipole moment (a) and polarizability components (b) of the test set consisting of 1200 points with respect to ML predictions of the models trained on 1000 points. Dipole moment components were learned with Mat\'{e}rn kernel, while for polarizability Gaussian kernel performed better. In both cases \texttt{RExyzCl} descriptor was used. Plots (c) and (d) are showing scalar and tensorial part  of the polarizability tensor.}
    \label{fig:scatter_plots}
\end{figure}

Geometries were taken after every 100 fs from this dataset, resulting in 1200-point test set for which reference dipole moments and polarizabilities were calculated using DFT.
The RMSE for this dataset was found to be four times larger than that of the test set constructed from the original training data. This justifies the construction of the new test set, rather than taking points from initial set of trajectories. 
Nevertheless, the ML predictions of dipole moment as well as polarizability components demonstrated a good agreement with the reference values, as shown on Fig.~\ref{fig:scatter_plots}, providing confidence in the model's predictive capabilities and the reliability of the proposed approach.

\begin{figure}
    \centering
    \includegraphics[height=.4\textwidth]{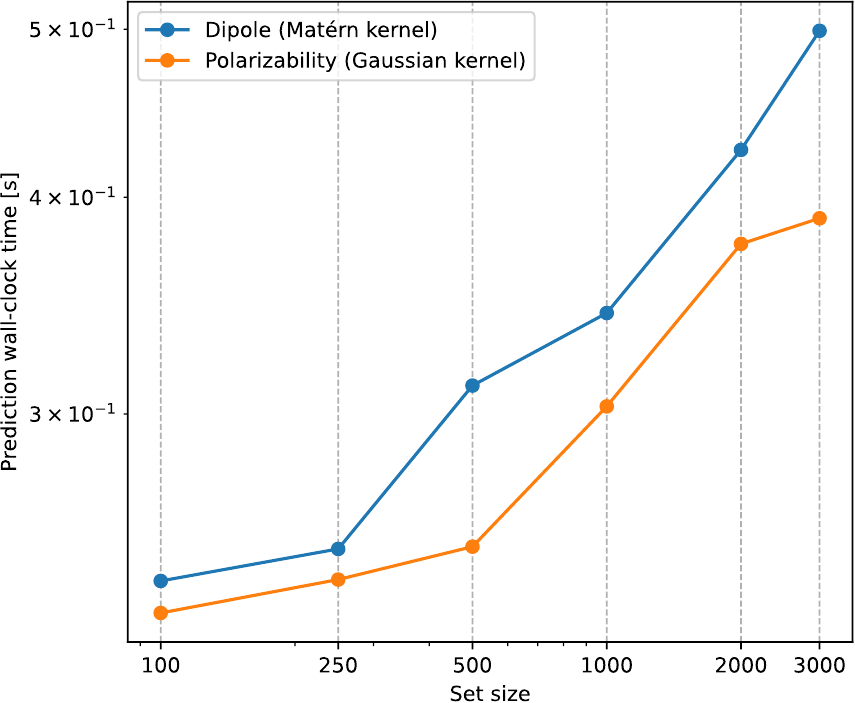}
    \includegraphics[height=.4\textwidth]{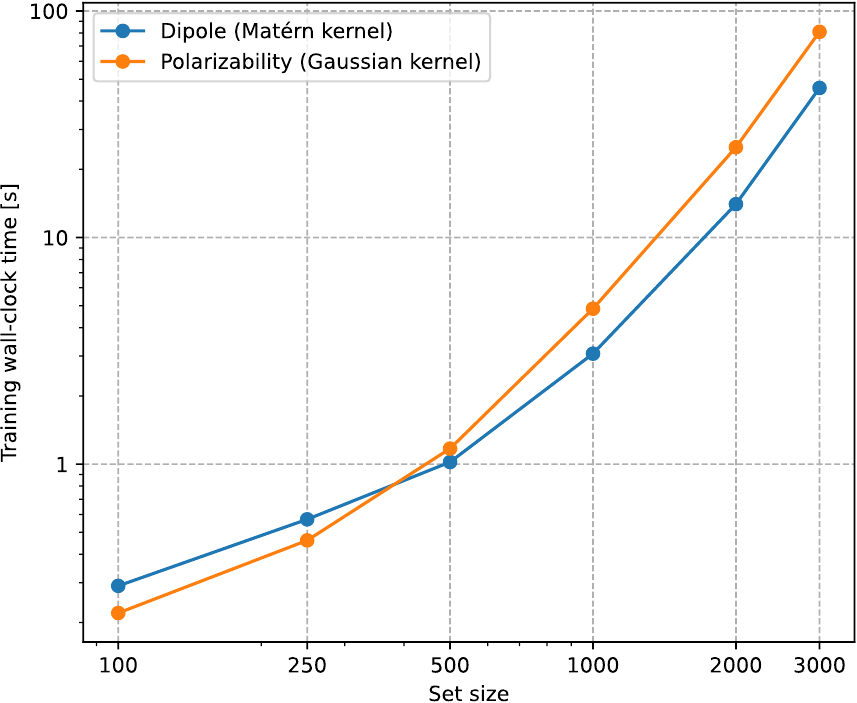}
    \caption{Average prediction and training times for dipole moments and polarizability tensor components as a function of data set size.}
    \label{fig:scale}
\end{figure}

To illustrate the efficiency of the presented procedure, Fig.~\ref{fig:scale} shows the times for the prediction and training 
as a function of data set size. The training and prediction times were estimated on a single core of an Intel(R) Xeon(R) Gold 6226R CPU running at 2.90~GHz. 
As can be seen, the training is extremely fast and takes less than 2~minutes for 3000~training points. 
The predictions are ca. a millisecond per point, which include the diagonalization of the inertia tensor and rotations of the molecular geometry, vectors and tensors that amount to only a fraction of the prediction time: 2~$\mu$s for one diagonalization and 0.08~$\mu$s for the rotation of one vector, thanks to the fact that only elementary linear algebra operations are involved
and no calculations of goniometric or special functions is required, unlike in some of the equivariant approaches. 
This high speed combined with the excellent accuracy makes the approach a practical choice for learning vectorial properties. 
If the single-core computations become too time consuming for larger molecules or datasets, a parallelized implementation with a good scaling is available in MLATOM \citep{MLatom2019}.
The RPR procedure is, e.g., particularly suitable for the transformation of a large number of vectors,
for example when NAD couplings are involved, where for a $N$-atom molecule and $M$ excited states there are $O(NM^2)$ NAD coupling vectors to be processed.

While the rotation procedure scales linearly with the number of atoms,
  the training/prediction calculations with our descriptor scale as $N(N-1)/2$ with the number of atoms $N$. 
  This is, e.g., faster than alternative approaches based on another global descriptor Coloumb matrix which scales as $N^2$. 
  We also know from the related benchmark of different machine learning potentials \citep{Pinheiro2021} that our RE descriptor with KRR provides the fastest (and, often, among the most accurate) solutions compared to methods with local descriptors such as SOAP or NN-based ANI, DeepPot, and particularly much faster than 'learned' descriptors such as in PhysNet and should be much faster than even slower equivariant descriptors.

\clearpage
\section{Conclusions}
In this study, we addressed the challenge of machine learning vector or tensor properties. We proposed rotate-predict-rotate, a simple three-step technique to guarantee proper covariance without the need for complex equivariant ML architectures or the construction of auxiliary properties. The RPR method involves rotating the molecular geometry to a "canonical" orientation defined by the eigenvectors of the molecular tensor of inertia, training the ML model on properties in this standard orientation, and finally rotating the predicted properties back to the original orientation. We used kernel ridge regression to fit dipole moment and polarizability components of 1,2-dichloroethane and evaluated its performance. Our results demonstrate that global molecular descriptors enhanced with cartesian coordinates of chlorine atoms significantly improves ML models accuracy.

The learning curves for dipole moment and polarizability components confirmed that a training set of 1000 points was sufficient to achieve satisfactory accuracy. We used the KREG model to propagate MD simulations and constructed a test set covering all possible spatial orientations and dihedral angles. This method provides a straightforward and computationally efficient solution for achieving rotational covariance in ML models for vector and tensor properties and is also extensible to higher-rank tensors.

\section*{Supplementary Material}
The supplementary material contains a Jupyter notebook and corresponding data related to descriptor benchmarks for energy, dipole moment, and polarizability components, along with other information required to generate the figures. This includes the evolution of dihedral angles, learning curves, scatter plots, predictions, and training times.

\section*{Acknowledgments}
The work of the Czech team has been supported by the \textit{Czech Science Foundation} Grant \mbox{23-06364S} and \textit{Charles University} (project GAUK 6224).
	We also highly appreciate the generous admission to computing
facilities owned by parties and projects contributing to the
National Grid Infrastructure MetaCentrum provided under the
program ``Projects of Large Infrastructure for Research,
Development, and Innovations ''(no. LM2010005) and computer time provided
by the IT4I supercomputing center (Project ID:90254) supported by the Ministry of Education, Youth and Sports. MB thanks the funding provided by the European Research Council (ERC) Advanced grant SubNano (Grant agreement 832237). MB received support from the French government under the France 2030, as part of the initiative d'Excellence d'Aix-Marseille Universit\'e, A*MIDEX (AMX-22-IN1-48). POD acknowledges funding by the National Natural Science Foundation of China (via the Outstanding Youth Scholars (Overseas, 2021) project).

\let\OLDthebibliography\thebibliography 
\renewcommand\thebibliography[1]{
\OLDthebibliography{#1}                                                                                 \setlength{\parskip}{0pt}
\setlength{\itemsep}{0pt plus 0.3ex}
}  

\bibliography{md,md2,ALL}

\end{document}

%% file: short.tex
\newcommand{\bI}{\mbox{\protect\boldmath $I$}}

\newcommand{\bK}{\mbox{\protect\boldmath $K$}}

\newcommand{\balpha}{\mbox{\protect\boldmath $\alpha$}}

\newcommand{\bx}{\mbox{\protect\boldmath $x$}}
\newcommand{\by}{\mbox{\protect\boldmath $y$}}

\newcommand{\br}{\mbox{\protect\boldmath $r$}}

\newcommand{\bQ}{\mbox{\protect\boldmath $Q$}}

\def\AmS{{\the\textfont2 A}\kern-.1667em\lower.5ex\hbox
     {\the\textfont2 M}\kern-.125em{\the\textfont2 S}}

\def\AW{Addison\kern.1em-\penalty0\hskip0pt Wesley}
\def\BibTeX{{\rm B\kern-.05em{\smc i\kern-.025emb}\kern-.08em\TeX}}




%% file: mlvector.bbl
\begin{thebibliography}{45}%
\makeatletter
\providecommand \@ifxundefined [1]{%
 \@ifx{#1\undefined}
}%
\providecommand \@ifnum [1]{%
 \ifnum #1\expandafter \@firstoftwo
 \else \expandafter \@secondoftwo
 \fi
}%
\providecommand \@ifx [1]{%
 \ifx #1\expandafter \@firstoftwo
 \else \expandafter \@secondoftwo
 \fi
}%
\providecommand \natexlab [1]{#1}%
\providecommand \enquote  [1]{``#1''}%
\providecommand \bibnamefont  [1]{#1}%
\providecommand \bibfnamefont [1]{#1}%
\providecommand \citenamefont [1]{#1}%
\providecommand \href@noop [0]{\@secondoftwo}%
\providecommand \href [0]{\begingroup \@sanitize@url \@href}%
\providecommand \@href[1]{\@@startlink{#1}\@@href}%
\providecommand \@@href[1]{\endgroup#1\@@endlink}%
\providecommand \@sanitize@url [0]{\catcode `\\12\catcode `\$12\catcode
  `\&12\catcode `\#12\catcode `\^12\catcode `\_12\catcode `\%12\relax}%
\providecommand \@@startlink[1]{}%
\providecommand \@@endlink[0]{}%
\providecommand \url  [0]{\begingroup\@sanitize@url \@url }%
\providecommand \@url [1]{\endgroup\@href {#1}{\urlprefix }}%
\providecommand \urlprefix  [0]{URL }%
\providecommand \Eprint [0]{\href }%
\providecommand \doibase [0]{http://dx.doi.org/}%
\providecommand \selectlanguage [0]{\@gobble}%
\providecommand \bibinfo  [0]{\@secondoftwo}%
\providecommand \bibfield  [0]{\@secondoftwo}%
\providecommand \translation [1]{[#1]}%
\providecommand \BibitemOpen [0]{}%
\providecommand \bibitemStop [0]{}%
\providecommand \bibitemNoStop [0]{.\EOS\space}%
\providecommand \EOS [0]{\spacefactor3000\relax}%
\providecommand \BibitemShut  [1]{\csname bibitem#1\endcsname}%
\let\auto@bib@innerbib\@empty
\bibitem [{\citenamefont {Dral}(2020)}]{Dral2020}%
  \BibitemOpen
  \bibfield  {author} {\bibinfo {author} {\bibfnamefont {P.~O.}\ \bibnamefont
  {Dral}},\ }\href {\doibase 10.1021/acs.jpclett.9b03664} {\bibfield  {journal}
  {\bibinfo  {journal} {The Journal of Physical Chemistry Letters}\ }\textbf
  {\bibinfo {volume} {11}},\ \bibinfo {pages} {2336} (\bibinfo {year}
  {2020})}\BibitemShut {NoStop}%
\bibitem [{\citenamefont {Lorenz}\ \emph {et~al.}(2004)\citenamefont {Lorenz},
  \citenamefont {Groß},\ and\ \citenamefont {Scheffler}}]{Lorenz2004}%
  \BibitemOpen
  \bibfield  {author} {\bibinfo {author} {\bibfnamefont {S.}~\bibnamefont
  {Lorenz}}, \bibinfo {author} {\bibfnamefont {A.}~\bibnamefont {Groß}}, \
  and\ \bibinfo {author} {\bibfnamefont {M.}~\bibnamefont {Scheffler}},\ }\href
  {\doibase 10.1016/j.cplett.2004.07.076} {\bibfield  {journal} {\bibinfo
  {journal} {Chemical Physics Letters}\ }\textbf {\bibinfo {volume} {395}},\
  \bibinfo {pages} {210} (\bibinfo {year} {2004})}\BibitemShut {NoStop}%
\bibitem [{\citenamefont {Raff}\ \emph {et~al.}(2005)\citenamefont {Raff},
  \citenamefont {Malshe}, \citenamefont {Hagan}, \citenamefont {Doughan},
  \citenamefont {Rockley},\ and\ \citenamefont {Komanduri}}]{Raff2005}%
  \BibitemOpen
  \bibfield  {author} {\bibinfo {author} {\bibfnamefont {L.~M.}\ \bibnamefont
  {Raff}}, \bibinfo {author} {\bibfnamefont {M.}~\bibnamefont {Malshe}},
  \bibinfo {author} {\bibfnamefont {M.}~\bibnamefont {Hagan}}, \bibinfo
  {author} {\bibfnamefont {D.~I.}\ \bibnamefont {Doughan}}, \bibinfo {author}
  {\bibfnamefont {M.~G.}\ \bibnamefont {Rockley}}, \ and\ \bibinfo {author}
  {\bibfnamefont {R.}~\bibnamefont {Komanduri}},\ }\href {\doibase
  10.1063/1.1850458} {\bibfield  {journal} {\bibinfo  {journal} {The Journal of
  Chemical Physics}\ }\textbf {\bibinfo {volume} {122}} (\bibinfo {year}
  {2005}),\ 10.1063/1.1850458}\BibitemShut {NoStop}%
\bibitem [{\citenamefont {Zhang}\ \emph {et~al.}(2023)\citenamefont {Zhang},
  \citenamefont {Jiang},\ and\ \citenamefont {Jiang}}]{Dral_book_19}%
  \BibitemOpen
  \bibfield  {author} {\bibinfo {author} {\bibfnamefont {Y.}~\bibnamefont
  {Zhang}}, \bibinfo {author} {\bibfnamefont {J.}~\bibnamefont {Jiang}}, \ and\
  \bibinfo {author} {\bibfnamefont {B.}~\bibnamefont {Jiang}},\ }in\ \href
  {\doibase https://doi.org/10.1016/B978-0-323-90049-2.00019-6} {\emph
  {\bibinfo {booktitle} {Quantum Chemistry in the Age of Machine Learning}}},\
  \bibinfo {editor} {edited by\ \bibinfo {editor} {\bibfnamefont {P.~O.}\
  \bibnamefont {Dral}}}\ (\bibinfo  {publisher} {Elsevier},\ \bibinfo {year}
  {2023})\ pp.\ \bibinfo {pages} {453--465}\BibitemShut {NoStop}%
\bibitem [{\citenamefont {Perakis}\ \emph {et~al.}(2016)\citenamefont
  {Perakis}, \citenamefont {De~Marco}, \citenamefont {Shalit}, \citenamefont
  {Tang}, \citenamefont {Kann}, \citenamefont {Kühne}, \citenamefont {Torre},
  \citenamefont {Bonn},\ and\ \citenamefont {Nagata}}]{Perakis2016}%
  \BibitemOpen
  \bibfield  {author} {\bibinfo {author} {\bibfnamefont {F.}~\bibnamefont
  {Perakis}}, \bibinfo {author} {\bibfnamefont {L.}~\bibnamefont {De~Marco}},
  \bibinfo {author} {\bibfnamefont {A.}~\bibnamefont {Shalit}}, \bibinfo
  {author} {\bibfnamefont {F.}~\bibnamefont {Tang}}, \bibinfo {author}
  {\bibfnamefont {Z.~R.}\ \bibnamefont {Kann}}, \bibinfo {author}
  {\bibfnamefont {T.~D.}\ \bibnamefont {Kühne}}, \bibinfo {author}
  {\bibfnamefont {R.}~\bibnamefont {Torre}}, \bibinfo {author} {\bibfnamefont
  {M.}~\bibnamefont {Bonn}}, \ and\ \bibinfo {author} {\bibfnamefont
  {Y.}~\bibnamefont {Nagata}},\ }\href {\doibase 10.1021/acs.chemrev.5b00640}
  {\bibfield  {journal} {\bibinfo  {journal} {Chemical Reviews}\ }\textbf
  {\bibinfo {volume} {116}},\ \bibinfo {pages} {7590} (\bibinfo {year}
  {2016})}\BibitemShut {NoStop}%
\bibitem [{\citenamefont {Zhang}\ \emph {et~al.}(2016)\citenamefont {Zhang},
  \citenamefont {Tan}, \citenamefont {Wu}, \citenamefont {Shi},\ and\
  \citenamefont {Tan}}]{Zhang2016}%
  \BibitemOpen
  \bibfield  {author} {\bibinfo {author} {\bibfnamefont {X.}~\bibnamefont
  {Zhang}}, \bibinfo {author} {\bibfnamefont {Q.-H.}\ \bibnamefont {Tan}},
  \bibinfo {author} {\bibfnamefont {J.-B.}\ \bibnamefont {Wu}}, \bibinfo
  {author} {\bibfnamefont {W.}~\bibnamefont {Shi}}, \ and\ \bibinfo {author}
  {\bibfnamefont {P.-H.}\ \bibnamefont {Tan}},\ }\href {\doibase
  10.1039/c5nr07205k} {\bibfield  {journal} {\bibinfo  {journal} {Nanoscale}\
  }\textbf {\bibinfo {volume} {8}},\ \bibinfo {pages} {6435} (\bibinfo {year}
  {2016})}\BibitemShut {NoStop}%
\bibitem [{\citenamefont {Cheng}\ \emph {et~al.}(2024)\citenamefont {Cheng},
  \citenamefont {Bi}, \citenamefont {Liu}, \citenamefont {Chen}, \citenamefont
  {Misquitta},\ and\ \citenamefont {Yu}}]{Cheng2024}%
  \BibitemOpen
  \bibfield  {author} {\bibinfo {author} {\bibfnamefont {Z.}~\bibnamefont
  {Cheng}}, \bibinfo {author} {\bibfnamefont {H.}~\bibnamefont {Bi}}, \bibinfo
  {author} {\bibfnamefont {S.}~\bibnamefont {Liu}}, \bibinfo {author}
  {\bibfnamefont {J.}~\bibnamefont {Chen}}, \bibinfo {author} {\bibfnamefont
  {A.~J.}\ \bibnamefont {Misquitta}}, \ and\ \bibinfo {author} {\bibfnamefont
  {K.}~\bibnamefont {Yu}},\ }\href {\doibase 10.1021/acs.jctc.4c00337}
  {\bibfield  {journal} {\bibinfo  {journal} {Journal of Chemical Theory and
  Computation}\ }\textbf {\bibinfo {volume} {20}},\ \bibinfo {pages} {5598}
  (\bibinfo {year} {2024})}\BibitemShut {NoStop}%
\bibitem [{\citenamefont {Ye}\ \emph {et~al.}(2019)\citenamefont {Ye},
  \citenamefont {Hu}, \citenamefont {Li}, \citenamefont {Zhang}, \citenamefont
  {Zhong}, \citenamefont {Zhang}, \citenamefont {Luo}, \citenamefont
  {Mukamel},\ and\ \citenamefont {Jiang}}]{Ye2019}%
  \BibitemOpen
  \bibfield  {author} {\bibinfo {author} {\bibfnamefont {S.}~\bibnamefont
  {Ye}}, \bibinfo {author} {\bibfnamefont {W.}~\bibnamefont {Hu}}, \bibinfo
  {author} {\bibfnamefont {X.}~\bibnamefont {Li}}, \bibinfo {author}
  {\bibfnamefont {J.}~\bibnamefont {Zhang}}, \bibinfo {author} {\bibfnamefont
  {K.}~\bibnamefont {Zhong}}, \bibinfo {author} {\bibfnamefont
  {G.}~\bibnamefont {Zhang}}, \bibinfo {author} {\bibfnamefont
  {Y.}~\bibnamefont {Luo}}, \bibinfo {author} {\bibfnamefont {S.}~\bibnamefont
  {Mukamel}}, \ and\ \bibinfo {author} {\bibfnamefont {J.}~\bibnamefont
  {Jiang}},\ }\href {\doibase 10.1073/pnas.1821044116} {\bibfield  {journal}
  {\bibinfo  {journal} {Proceedings of the National Academy of Sciences}\
  }\textbf {\bibinfo {volume} {116}},\ \bibinfo {pages} {11612} (\bibinfo
  {year} {2019})}\BibitemShut {NoStop}%
\bibitem [{\citenamefont {Zhang}\ \emph {et~al.}(2021)\citenamefont {Zhang},
  \citenamefont {Ye}, \citenamefont {Zhong}, \citenamefont {Zhang},
  \citenamefont {Chong}, \citenamefont {Zhao}, \citenamefont {Zhou},
  \citenamefont {Guo}, \citenamefont {Zhang}, \citenamefont {Jiang},
  \citenamefont {Mukamel},\ and\ \citenamefont {Jiang}}]{Zhang2021}%
  \BibitemOpen
  \bibfield  {author} {\bibinfo {author} {\bibfnamefont {J.}~\bibnamefont
  {Zhang}}, \bibinfo {author} {\bibfnamefont {S.}~\bibnamefont {Ye}}, \bibinfo
  {author} {\bibfnamefont {K.}~\bibnamefont {Zhong}}, \bibinfo {author}
  {\bibfnamefont {Y.}~\bibnamefont {Zhang}}, \bibinfo {author} {\bibfnamefont
  {Y.}~\bibnamefont {Chong}}, \bibinfo {author} {\bibfnamefont
  {L.}~\bibnamefont {Zhao}}, \bibinfo {author} {\bibfnamefont {H.}~\bibnamefont
  {Zhou}}, \bibinfo {author} {\bibfnamefont {S.}~\bibnamefont {Guo}}, \bibinfo
  {author} {\bibfnamefont {G.}~\bibnamefont {Zhang}}, \bibinfo {author}
  {\bibfnamefont {B.}~\bibnamefont {Jiang}}, \bibinfo {author} {\bibfnamefont
  {S.}~\bibnamefont {Mukamel}}, \ and\ \bibinfo {author} {\bibfnamefont
  {J.}~\bibnamefont {Jiang}},\ }\href {\doibase 10.1021/acs.jpcb.1c03296}
  {\bibfield  {journal} {\bibinfo  {journal} {The Journal of Physical Chemistry
  B}\ }\textbf {\bibinfo {volume} {125}},\ \bibinfo {pages} {6171} (\bibinfo
  {year} {2021})}\BibitemShut {NoStop}%
\bibitem [{\citenamefont {Ye}\ \emph {et~al.}(2020)\citenamefont {Ye},
  \citenamefont {Zhong}, \citenamefont {Zhang}, \citenamefont {Hu},
  \citenamefont {Hirst}, \citenamefont {Zhang}, \citenamefont {Mukamel},\ and\
  \citenamefont {Jiang}}]{Ye2020}%
  \BibitemOpen
  \bibfield  {author} {\bibinfo {author} {\bibfnamefont {S.}~\bibnamefont
  {Ye}}, \bibinfo {author} {\bibfnamefont {K.}~\bibnamefont {Zhong}}, \bibinfo
  {author} {\bibfnamefont {J.}~\bibnamefont {Zhang}}, \bibinfo {author}
  {\bibfnamefont {W.}~\bibnamefont {Hu}}, \bibinfo {author} {\bibfnamefont
  {J.~D.}\ \bibnamefont {Hirst}}, \bibinfo {author} {\bibfnamefont
  {G.}~\bibnamefont {Zhang}}, \bibinfo {author} {\bibfnamefont
  {S.}~\bibnamefont {Mukamel}}, \ and\ \bibinfo {author} {\bibfnamefont
  {J.}~\bibnamefont {Jiang}},\ }\href {\doibase 10.1021/jacs.0c06530}
  {\bibfield  {journal} {\bibinfo  {journal} {Journal of the American Chemical
  Society}\ }\textbf {\bibinfo {volume} {142}},\ \bibinfo {pages} {19071}
  (\bibinfo {year} {2020})}\BibitemShut {NoStop}%
\bibitem [{\citenamefont {Westermayr}\ \emph {et~al.}(2019)\citenamefont
  {Westermayr}, \citenamefont {Gastegger}, \citenamefont {Menger},
  \citenamefont {Mai}, \citenamefont {Gonz{\'{a}}lez},\ and\ \citenamefont
  {Marquetand}}]{Westermayr2019}%
  \BibitemOpen
  \bibfield  {author} {\bibinfo {author} {\bibfnamefont {J.}~\bibnamefont
  {Westermayr}}, \bibinfo {author} {\bibfnamefont {M.}~\bibnamefont
  {Gastegger}}, \bibinfo {author} {\bibfnamefont {M.~F. S.~J.}\ \bibnamefont
  {Menger}}, \bibinfo {author} {\bibfnamefont {S.}~\bibnamefont {Mai}},
  \bibinfo {author} {\bibfnamefont {L.}~\bibnamefont {Gonz{\'{a}}lez}}, \ and\
  \bibinfo {author} {\bibfnamefont {P.}~\bibnamefont {Marquetand}},\ }\href
  {\doibase 10.1039/c9sc01742a} {\bibfield  {journal} {\bibinfo  {journal}
  {Chemical Science}\ }\textbf {\bibinfo {volume} {10}},\ \bibinfo {pages}
  {8100} (\bibinfo {year} {2019})}\BibitemShut {NoStop}%
\bibitem [{\citenamefont {Westermayr}\ \emph
  {et~al.}(2020{\natexlab{a}})\citenamefont {Westermayr}, \citenamefont
  {Faber}, \citenamefont {Christensen}, \citenamefont {von Lilienfeld},\ and\
  \citenamefont {Marquetand}}]{Westermayr2020a}%
  \BibitemOpen
  \bibfield  {author} {\bibinfo {author} {\bibfnamefont {J.}~\bibnamefont
  {Westermayr}}, \bibinfo {author} {\bibfnamefont {F.~A.}\ \bibnamefont
  {Faber}}, \bibinfo {author} {\bibfnamefont {A.~S.}\ \bibnamefont
  {Christensen}}, \bibinfo {author} {\bibfnamefont {O.~A.}\ \bibnamefont {von
  Lilienfeld}}, \ and\ \bibinfo {author} {\bibfnamefont {P.}~\bibnamefont
  {Marquetand}},\ }\href {\doibase 10.1088/2632-2153/ab88d0} {\bibfield
  {journal} {\bibinfo  {journal} {Machine Learning: Science and Technology}\
  }\textbf {\bibinfo {volume} {1}},\ \bibinfo {pages} {025009} (\bibinfo {year}
  {2020}{\natexlab{a}})}\BibitemShut {NoStop}%
\bibitem [{\citenamefont {Westermayr}\ \emph
  {et~al.}(2020{\natexlab{b}})\citenamefont {Westermayr}, \citenamefont
  {Gastegger},\ and\ \citenamefont {Marquetand}}]{Westermayr2020b}%
  \BibitemOpen
  \bibfield  {author} {\bibinfo {author} {\bibfnamefont {J.}~\bibnamefont
  {Westermayr}}, \bibinfo {author} {\bibfnamefont {M.}~\bibnamefont
  {Gastegger}}, \ and\ \bibinfo {author} {\bibfnamefont {P.}~\bibnamefont
  {Marquetand}},\ }\href {\doibase 10.1021/acs.jpclett.0c00527} {\bibfield
  {journal} {\bibinfo  {journal} {The Journal of Physical Chemistry Letters}\
  }\textbf {\bibinfo {volume} {11}},\ \bibinfo {pages} {3828} (\bibinfo {year}
  {2020}{\natexlab{b}})}\BibitemShut {NoStop}%
\bibitem [{\citenamefont {Richardson}(2023)}]{Richardson2023}%
  \BibitemOpen
  \bibfield  {author} {\bibinfo {author} {\bibfnamefont {J.~O.}\ \bibnamefont
  {Richardson}},\ }\href {\doibase 10.1063/5.0133191} {\bibfield  {journal}
  {\bibinfo  {journal} {The Journal of Chemical Physics}\ }\textbf {\bibinfo
  {volume} {158}},\ \bibinfo {pages} {011102} (\bibinfo {year}
  {2023})}\BibitemShut {NoStop}%
\bibitem [{\citenamefont {Unke}\ and\ \citenamefont
  {Meuwly}(2019)}]{PhysNet2019}%
  \BibitemOpen
  \bibfield  {author} {\bibinfo {author} {\bibfnamefont {O.~T.}\ \bibnamefont
  {Unke}}\ and\ \bibinfo {author} {\bibfnamefont {M.}~\bibnamefont {Meuwly}},\
  }\href {\doibase 10.1021/acs.jctc.9b00181} {\bibfield  {journal} {\bibinfo
  {journal} {J. Chem. Theory Comput.}\ }\textbf {\bibinfo {volume} {15}},\
  \bibinfo {pages} {3678–} (\bibinfo {year} {2019})}\BibitemShut {NoStop}%
\bibitem [{\citenamefont {Chen}\ \emph
  {et~al.}(2024{\natexlab{a}})\citenamefont {Chen}, \citenamefont {Pios},
  \citenamefont {Gelin},\ and\ \citenamefont {Chen}}]{VSpecNN2024}%
  \BibitemOpen
  \bibfield  {author} {\bibinfo {author} {\bibfnamefont {Y.}~\bibnamefont
  {Chen}}, \bibinfo {author} {\bibfnamefont {S.~V.}\ \bibnamefont {Pios}},
  \bibinfo {author} {\bibfnamefont {M.~F.}\ \bibnamefont {Gelin}}, \ and\
  \bibinfo {author} {\bibfnamefont {L.}~\bibnamefont {Chen}},\ }\href {\doibase
  https://doi.org/10.1021/acs.jctc.4c00173} {\bibfield  {journal} {\bibinfo
  {journal} {Journal of Chemical Theory and Computation}\ }\textbf {\bibinfo
  {volume} {20}},\ \bibinfo {pages} {4703} (\bibinfo {year}
  {2024}{\natexlab{a}})}\BibitemShut {NoStop}%
\bibitem [{\citenamefont {Zhang}\ \emph
  {et~al.}(2020{\natexlab{a}})\citenamefont {Zhang}, \citenamefont {Ye},
  \citenamefont {Zhang}, \citenamefont {Hu}, \citenamefont {Jiang},\ and\
  \citenamefont {Jiang}}]{Zhang2020}%
  \BibitemOpen
  \bibfield  {author} {\bibinfo {author} {\bibfnamefont {Y.}~\bibnamefont
  {Zhang}}, \bibinfo {author} {\bibfnamefont {S.}~\bibnamefont {Ye}}, \bibinfo
  {author} {\bibfnamefont {J.}~\bibnamefont {Zhang}}, \bibinfo {author}
  {\bibfnamefont {C.}~\bibnamefont {Hu}}, \bibinfo {author} {\bibfnamefont
  {J.}~\bibnamefont {Jiang}}, \ and\ \bibinfo {author} {\bibfnamefont
  {B.}~\bibnamefont {Jiang}},\ }\href {\doibase 10.1021/acs.jpcb.0c06926}
  {\bibfield  {journal} {\bibinfo  {journal} {The Journal of Physical Chemistry
  B}\ }\textbf {\bibinfo {volume} {124}},\ \bibinfo {pages} {7284} (\bibinfo
  {year} {2020}{\natexlab{a}})}\BibitemShut {NoStop}%
\bibitem [{\citenamefont {Christensen}\ \emph {et~al.}(2019)\citenamefont
  {Christensen}, \citenamefont {Faber},\ and\ \citenamefont {von
  Lilienfeld}}]{christensen2019}%
  \BibitemOpen
  \bibfield  {author} {\bibinfo {author} {\bibfnamefont {A.~S.}\ \bibnamefont
  {Christensen}}, \bibinfo {author} {\bibfnamefont {F.~A.}\ \bibnamefont
  {Faber}}, \ and\ \bibinfo {author} {\bibfnamefont {O.~A.}\ \bibnamefont {von
  Lilienfeld}},\ }\href@noop {} {\bibfield  {journal} {\bibinfo  {journal} {J.
  Chem. Phys.}\ }\textbf {\bibinfo {volume} {150}},\ \bibinfo {pages} {064105}
  (\bibinfo {year} {2019})}\BibitemShut {NoStop}%
\bibitem [{\citenamefont {Bartók}\ \emph {et~al.}(2013)\citenamefont
  {Bartók}, \citenamefont {Kondor},\ and\ \citenamefont
  {Csányi}}]{Bartok2013}%
  \BibitemOpen
  \bibfield  {author} {\bibinfo {author} {\bibfnamefont {A.~P.}\ \bibnamefont
  {Bartók}}, \bibinfo {author} {\bibfnamefont {R.}~\bibnamefont {Kondor}}, \
  and\ \bibinfo {author} {\bibfnamefont {G.}~\bibnamefont {Csányi}},\ }\href
  {\doibase 10.1103/physrevb.87.184115} {\bibfield  {journal} {\bibinfo
  {journal} {Physical Review B}\ }\textbf {\bibinfo {volume} {87}},\ \bibinfo
  {pages} {184115} (\bibinfo {year} {2013})}\BibitemShut {NoStop}%
\bibitem [{\citenamefont {Glielmo}\ \emph {et~al.}(2017)\citenamefont
  {Glielmo}, \citenamefont {Sollich},\ and\ \citenamefont
  {De~Vita}}]{Glielmo2017}%
  \BibitemOpen
  \bibfield  {author} {\bibinfo {author} {\bibfnamefont {A.}~\bibnamefont
  {Glielmo}}, \bibinfo {author} {\bibfnamefont {P.}~\bibnamefont {Sollich}}, \
  and\ \bibinfo {author} {\bibfnamefont {A.}~\bibnamefont {De~Vita}},\ }\href
  {\doibase 10.1103/physrevb.95.214302} {\bibfield  {journal} {\bibinfo
  {journal} {Physical Review B}\ }\textbf {\bibinfo {volume} {95}},\ \bibinfo
  {pages} {214302} (\bibinfo {year} {2017})}\BibitemShut {NoStop}%
\bibitem [{\citenamefont {Grisafi}\ \emph {et~al.}(2018)\citenamefont
  {Grisafi}, \citenamefont {Wilkins}, \citenamefont {Cs{\'{a}}nyi},\ and\
  \citenamefont {Ceriotti}}]{Grisafi2018}%
  \BibitemOpen
  \bibfield  {author} {\bibinfo {author} {\bibfnamefont {A.}~\bibnamefont
  {Grisafi}}, \bibinfo {author} {\bibfnamefont {D.~M.}\ \bibnamefont
  {Wilkins}}, \bibinfo {author} {\bibfnamefont {G.}~\bibnamefont
  {Cs{\'{a}}nyi}}, \ and\ \bibinfo {author} {\bibfnamefont {M.}~\bibnamefont
  {Ceriotti}},\ }\href {\doibase 10.1103/physrevlett.120.036002} {\bibfield
  {journal} {\bibinfo  {journal} {Physical Review Letters}\ }\textbf {\bibinfo
  {volume} {120}},\ \bibinfo {pages} {036002} (\bibinfo {year}
  {2018})}\BibitemShut {NoStop}%
\bibitem [{\citenamefont {Wilkins}\ \emph {et~al.}(2019)\citenamefont
  {Wilkins}, \citenamefont {Grisafi}, \citenamefont {Yang}, \citenamefont
  {Lao}, \citenamefont {DiStasio},\ and\ \citenamefont
  {Ceriotti}}]{Wilkins2019}%
  \BibitemOpen
  \bibfield  {author} {\bibinfo {author} {\bibfnamefont {D.~M.}\ \bibnamefont
  {Wilkins}}, \bibinfo {author} {\bibfnamefont {A.}~\bibnamefont {Grisafi}},
  \bibinfo {author} {\bibfnamefont {Y.}~\bibnamefont {Yang}}, \bibinfo {author}
  {\bibfnamefont {K.~U.}\ \bibnamefont {Lao}}, \bibinfo {author} {\bibfnamefont
  {R.~A.}\ \bibnamefont {DiStasio}}, \ and\ \bibinfo {author} {\bibfnamefont
  {M.}~\bibnamefont {Ceriotti}},\ }\href {\doibase 10.1073/pnas.1816132116}
  {\bibfield  {journal} {\bibinfo  {journal} {Proceedings of the National
  Academy of Sciences}\ }\textbf {\bibinfo {volume} {116}},\ \bibinfo {pages}
  {3401} (\bibinfo {year} {2019})}\BibitemShut {NoStop}%
\bibitem [{\citenamefont {Raimbault}\ \emph {et~al.}(2019)\citenamefont
  {Raimbault}, \citenamefont {Grisafi}, \citenamefont {Ceriotti},\ and\
  \citenamefont {Rossi}}]{Raimbault2019}%
  \BibitemOpen
  \bibfield  {author} {\bibinfo {author} {\bibfnamefont {N.}~\bibnamefont
  {Raimbault}}, \bibinfo {author} {\bibfnamefont {A.}~\bibnamefont {Grisafi}},
  \bibinfo {author} {\bibfnamefont {M.}~\bibnamefont {Ceriotti}}, \ and\
  \bibinfo {author} {\bibfnamefont {M.}~\bibnamefont {Rossi}},\ }\href
  {\doibase 10.1088/1367-2630/ab4509} {\bibfield  {journal} {\bibinfo
  {journal} {New Journal of Physics}\ }\textbf {\bibinfo {volume} {21}},\
  \bibinfo {pages} {105001} (\bibinfo {year} {2019})}\BibitemShut {NoStop}%
\bibitem [{\citenamefont {Fan}\ \emph {et~al.}(2021)\citenamefont {Fan},
  \citenamefont {Zeng}, \citenamefont {Zhang}, \citenamefont {Wang},
  \citenamefont {Song}, \citenamefont {Dong}, \citenamefont {Chen},\ and\
  \citenamefont {Ala-Nissila}}]{Fan2021}%
  \BibitemOpen
  \bibfield  {author} {\bibinfo {author} {\bibfnamefont {Z.}~\bibnamefont
  {Fan}}, \bibinfo {author} {\bibfnamefont {Z.}~\bibnamefont {Zeng}}, \bibinfo
  {author} {\bibfnamefont {C.}~\bibnamefont {Zhang}}, \bibinfo {author}
  {\bibfnamefont {Y.}~\bibnamefont {Wang}}, \bibinfo {author} {\bibfnamefont
  {K.}~\bibnamefont {Song}}, \bibinfo {author} {\bibfnamefont {H.}~\bibnamefont
  {Dong}}, \bibinfo {author} {\bibfnamefont {Y.}~\bibnamefont {Chen}}, \ and\
  \bibinfo {author} {\bibfnamefont {T.}~\bibnamefont {Ala-Nissila}},\ }\href
  {\doibase 10.1103/physrevb.104.104309} {\bibfield  {journal} {\bibinfo
  {journal} {Physical Review B}\ }\textbf {\bibinfo {volume} {104}},\ \bibinfo
  {pages} {104309} (\bibinfo {year} {2021})}\BibitemShut {NoStop}%
\bibitem [{\citenamefont {Xu}\ \emph {et~al.}(2024)\citenamefont {Xu},
  \citenamefont {Rosander}, \citenamefont {Schäfer}, \citenamefont {Lindgren},
  \citenamefont {Österbacka}, \citenamefont {Fang}, \citenamefont {Chen},
  \citenamefont {He}, \citenamefont {Fan},\ and\ \citenamefont
  {Erhart}}]{Xu2024}%
  \BibitemOpen
  \bibfield  {author} {\bibinfo {author} {\bibfnamefont {N.}~\bibnamefont
  {Xu}}, \bibinfo {author} {\bibfnamefont {P.}~\bibnamefont {Rosander}},
  \bibinfo {author} {\bibfnamefont {C.}~\bibnamefont {Schäfer}}, \bibinfo
  {author} {\bibfnamefont {E.}~\bibnamefont {Lindgren}}, \bibinfo {author}
  {\bibfnamefont {N.}~\bibnamefont {Österbacka}}, \bibinfo {author}
  {\bibfnamefont {M.}~\bibnamefont {Fang}}, \bibinfo {author} {\bibfnamefont
  {W.}~\bibnamefont {Chen}}, \bibinfo {author} {\bibfnamefont {Y.}~\bibnamefont
  {He}}, \bibinfo {author} {\bibfnamefont {Z.}~\bibnamefont {Fan}}, \ and\
  \bibinfo {author} {\bibfnamefont {P.}~\bibnamefont {Erhart}},\ }\href
  {\doibase 10.1021/acs.jctc.3c01343} {\bibfield  {journal} {\bibinfo
  {journal} {Journal of Chemical Theory and Computation}\ }\textbf {\bibinfo
  {volume} {20}},\ \bibinfo {pages} {3273} (\bibinfo {year}
  {2024})}\BibitemShut {NoStop}%
\bibitem [{\citenamefont {Zhang}\ \emph {et~al.}(2019)\citenamefont {Zhang},
  \citenamefont {Hu},\ and\ \citenamefont {Jiang}}]{Zhang2019}%
  \BibitemOpen
  \bibfield  {author} {\bibinfo {author} {\bibfnamefont {Y.}~\bibnamefont
  {Zhang}}, \bibinfo {author} {\bibfnamefont {C.}~\bibnamefont {Hu}}, \ and\
  \bibinfo {author} {\bibfnamefont {B.}~\bibnamefont {Jiang}},\ }\href
  {\doibase 10.1021/acs.jpclett.9b02037} {\bibfield  {journal} {\bibinfo
  {journal} {The Journal of Physical Chemistry Letters}\ }\textbf {\bibinfo
  {volume} {10}},\ \bibinfo {pages} {4962} (\bibinfo {year}
  {2019})}\BibitemShut {NoStop}%
\bibitem [{\citenamefont {Anderson}\ \emph {et~al.}(2019)\citenamefont
  {Anderson}, \citenamefont {Hy},\ and\ \citenamefont {Kondor}}]{Anderson2019}%
  \BibitemOpen
  \bibfield  {author} {\bibinfo {author} {\bibfnamefont {B.}~\bibnamefont
  {Anderson}}, \bibinfo {author} {\bibfnamefont {T.~S.}\ \bibnamefont {Hy}}, \
  and\ \bibinfo {author} {\bibfnamefont {R.}~\bibnamefont {Kondor}},\ }in\
  \href
  {https://proceedings.neurips.cc/paper_files/paper/2019/file/03573b32b2746e6e8ca98b9123f2249b-Paper.pdf}
  {\emph {\bibinfo {booktitle} {Advances in Neural Information Processing
  Systems}}},\ Vol.~\bibinfo {volume} {32},\ \bibinfo {editor} {edited by\
  \bibinfo {editor} {\bibfnamefont {H.}~\bibnamefont {Wallach}}, \bibinfo
  {editor} {\bibfnamefont {H.}~\bibnamefont {Larochelle}}, \bibinfo {editor}
  {\bibfnamefont {A.}~\bibnamefont {Beygelzimer}}, \bibinfo {editor}
  {\bibfnamefont {F.}~\bibnamefont {d\textquotesingle Alch\'{e}-Buc}}, \bibinfo
  {editor} {\bibfnamefont {E.}~\bibnamefont {Fox}}, \ and\ \bibinfo {editor}
  {\bibfnamefont {R.}~\bibnamefont {Garnett}}}\ (\bibinfo  {publisher} {Curran
  Associates, Inc.},\ \bibinfo {year} {2019})\BibitemShut {NoStop}%
\bibitem [{\citenamefont {Sch{\"u}tt}\ \emph {et~al.}(2021)\citenamefont
  {Sch{\"u}tt}, \citenamefont {Unke},\ and\ \citenamefont
  {Gastegger}}]{Schutt2021}%
  \BibitemOpen
  \bibfield  {author} {\bibinfo {author} {\bibfnamefont {K.~T.}\ \bibnamefont
  {Sch{\"u}tt}}, \bibinfo {author} {\bibfnamefont {O.~T.}\ \bibnamefont
  {Unke}}, \ and\ \bibinfo {author} {\bibfnamefont {M.}~\bibnamefont
  {Gastegger}},\ }in\ \href
  {https://api.semanticscholar.org/CorpusID:231839844} {\emph {\bibinfo
  {booktitle} {International Conference on Machine Learning}}}\ (\bibinfo
  {year} {2021})\BibitemShut {NoStop}%
\bibitem [{\citenamefont {Thomas}\ \emph {et~al.}(2018)\citenamefont {Thomas},
  \citenamefont {Smidt}, \citenamefont {Kearnes}, \citenamefont {Yang},
  \citenamefont {Li}, \citenamefont {Kohlhoff},\ and\ \citenamefont
  {Riley}}]{Thomas2018}%
  \BibitemOpen
  \bibfield  {author} {\bibinfo {author} {\bibfnamefont {N.}~\bibnamefont
  {Thomas}}, \bibinfo {author} {\bibfnamefont {T.}~\bibnamefont {Smidt}},
  \bibinfo {author} {\bibfnamefont {S.}~\bibnamefont {Kearnes}}, \bibinfo
  {author} {\bibfnamefont {L.}~\bibnamefont {Yang}}, \bibinfo {author}
  {\bibfnamefont {L.}~\bibnamefont {Li}}, \bibinfo {author} {\bibfnamefont
  {K.}~\bibnamefont {Kohlhoff}}, \ and\ \bibinfo {author} {\bibfnamefont
  {P.}~\bibnamefont {Riley}},\ }\href {\doibase 10.48550/ARXIV.1802.08219}
  {\enquote {\bibinfo {title} {Tensor field networks: Rotation- and
  translation-equivariant neural networks for 3d point clouds},}\ } (\bibinfo
  {year} {2018}),\ \bibinfo {note} {arXiv:1802.08219}\BibitemShut {NoStop}%
\bibitem [{\citenamefont {Chen}\ \emph
  {et~al.}(2024{\natexlab{b}})\citenamefont {Chen}, \citenamefont {Pios},
  \citenamefont {Gelin},\ and\ \citenamefont {Chen}}]{Chen2024}%
  \BibitemOpen
  \bibfield  {author} {\bibinfo {author} {\bibfnamefont {Y.}~\bibnamefont
  {Chen}}, \bibinfo {author} {\bibfnamefont {S.~V.}\ \bibnamefont {Pios}},
  \bibinfo {author} {\bibfnamefont {M.~F.}\ \bibnamefont {Gelin}}, \ and\
  \bibinfo {author} {\bibfnamefont {L.}~\bibnamefont {Chen}},\ }\href {\doibase
  10.1021/acs.jctc.4c00173} {\bibfield  {journal} {\bibinfo  {journal} {Journal
  of Chemical Theory and Computation}\ }\textbf {\bibinfo {volume} {20}},\
  \bibinfo {pages} {4703} (\bibinfo {year} {2024}{\natexlab{b}})}\BibitemShut
  {NoStop}%
\bibitem [{\citenamefont {Zhang}\ \emph
  {et~al.}(2020{\natexlab{b}})\citenamefont {Zhang}, \citenamefont {Chen},
  \citenamefont {Wu}, \citenamefont {Wang}, \citenamefont {E},\ and\
  \citenamefont {Car}}]{Zhang2020a}%
  \BibitemOpen
  \bibfield  {author} {\bibinfo {author} {\bibfnamefont {L.}~\bibnamefont
  {Zhang}}, \bibinfo {author} {\bibfnamefont {M.}~\bibnamefont {Chen}},
  \bibinfo {author} {\bibfnamefont {X.}~\bibnamefont {Wu}}, \bibinfo {author}
  {\bibfnamefont {H.}~\bibnamefont {Wang}}, \bibinfo {author} {\bibfnamefont
  {W.}~\bibnamefont {E}}, \ and\ \bibinfo {author} {\bibfnamefont
  {R.}~\bibnamefont {Car}},\ }\href {\doibase 10.1103/physrevb.102.041121}
  {\bibfield  {journal} {\bibinfo  {journal} {Physical Review B}\ }\textbf
  {\bibinfo {volume} {102}},\ \bibinfo {pages} {041121} (\bibinfo {year}
  {2020}{\natexlab{b}})}\BibitemShut {NoStop}%
\bibitem [{\citenamefont {Zhang}\ and\ \citenamefont
  {Jiang}(2023)}]{Zhang2023a}%
  \BibitemOpen
  \bibfield  {author} {\bibinfo {author} {\bibfnamefont {Y.}~\bibnamefont
  {Zhang}}\ and\ \bibinfo {author} {\bibfnamefont {B.}~\bibnamefont {Jiang}},\
  }\href {\doibase 10.1038/s41467-023-42148-y} {\bibfield  {journal} {\bibinfo
  {journal} {Nature Communications}\ }\textbf {\bibinfo {volume} {14}}
  (\bibinfo {year} {2023}),\ 10.1038/s41467-023-42148-y}\BibitemShut {NoStop}%
\bibitem [{\citenamefont {Bereau}\ \emph
  {et~al.}(2015{\natexlab{a}})\citenamefont {Bereau}, \citenamefont
  {Andrienko},\ and\ \citenamefont {von Lilienfeld}}]{bereau2015b}%
  \BibitemOpen
  \bibfield  {author} {\bibinfo {author} {\bibfnamefont {T.}~\bibnamefont
  {Bereau}}, \bibinfo {author} {\bibfnamefont {D.}~\bibnamefont {Andrienko}}, \
  and\ \bibinfo {author} {\bibfnamefont {O.~A.}\ \bibnamefont {von
  Lilienfeld}},\ }\href@noop {} {\bibfield  {journal} {\bibinfo  {journal} {J.
  Chem. Theory Comp.}\ }\textbf {\bibinfo {volume} {11}},\ \bibinfo {pages}
  {3225} (\bibinfo {year} {2015}{\natexlab{a}})}\BibitemShut {NoStop}%
\bibitem [{\citenamefont {Rupp}\ \emph {et~al.}(2015)\citenamefont {Rupp},
  \citenamefont {Ramakrishnan},\ and\ \citenamefont {von
  Lilienfeld}}]{rupp2015}%
  \BibitemOpen
  \bibfield  {author} {\bibinfo {author} {\bibfnamefont {M.}~\bibnamefont
  {Rupp}}, \bibinfo {author} {\bibfnamefont {R.}~\bibnamefont {Ramakrishnan}},
  \ and\ \bibinfo {author} {\bibfnamefont {O.~A.}\ \bibnamefont {von
  Lilienfeld}},\ }\href@noop {} {\bibfield  {journal} {\bibinfo  {journal} {J.
  Phys. Chem. Lett.}\ }\textbf {\bibinfo {volume} {6}},\ \bibinfo {pages}
  {3309} (\bibinfo {year} {2015})}\BibitemShut {NoStop}%
\bibitem [{\citenamefont {Bereau}\ \emph
  {et~al.}(2015{\natexlab{b}})\citenamefont {Bereau}, \citenamefont
  {Andrienko},\ and\ \citenamefont {von Lilienfeld}}]{Bereau2015}%
  \BibitemOpen
  \bibfield  {author} {\bibinfo {author} {\bibfnamefont {T.}~\bibnamefont
  {Bereau}}, \bibinfo {author} {\bibfnamefont {D.}~\bibnamefont {Andrienko}}, \
  and\ \bibinfo {author} {\bibfnamefont {O.~A.}\ \bibnamefont {von
  Lilienfeld}},\ }\href {\doibase 10.1021/acs.jctc.5b00301} {\bibfield
  {journal} {\bibinfo  {journal} {Journal of Chemical Theory and Computation}\
  }\textbf {\bibinfo {volume} {11}},\ \bibinfo {pages} {3225} (\bibinfo {year}
  {2015}{\natexlab{b}})}\BibitemShut {NoStop}%
\bibitem [{\citenamefont {Liang}\ \emph {et~al.}(2017)\citenamefont {Liang},
  \citenamefont {Tocci}, \citenamefont {Wilkins}, \citenamefont {Grisafi},
  \citenamefont {Roke},\ and\ \citenamefont {Ceriotti}}]{Liang2017}%
  \BibitemOpen
  \bibfield  {author} {\bibinfo {author} {\bibfnamefont {C.}~\bibnamefont
  {Liang}}, \bibinfo {author} {\bibfnamefont {G.}~\bibnamefont {Tocci}},
  \bibinfo {author} {\bibfnamefont {D.~M.}\ \bibnamefont {Wilkins}}, \bibinfo
  {author} {\bibfnamefont {A.}~\bibnamefont {Grisafi}}, \bibinfo {author}
  {\bibfnamefont {S.}~\bibnamefont {Roke}}, \ and\ \bibinfo {author}
  {\bibfnamefont {M.}~\bibnamefont {Ceriotti}},\ }\href {\doibase
  10.1103/physrevb.96.041407} {\bibfield  {journal} {\bibinfo  {journal}
  {Physical Review B}\ }\textbf {\bibinfo {volume} {96}},\ \bibinfo {pages}
  {041407} (\bibinfo {year} {2017})}\BibitemShut {NoStop}%
\bibitem [{\citenamefont {Hu}\ and\ \citenamefont {Huo}(2023)}]{HuHuo2023}%
  \BibitemOpen
  \bibfield  {author} {\bibinfo {author} {\bibfnamefont {D.}~\bibnamefont
  {Hu}}\ and\ \bibinfo {author} {\bibfnamefont {P.}~\bibnamefont {Huo}},\
  }\href@noop {} {\bibfield  {journal} {\bibinfo  {journal} {J. Chem. Theory
  Comp.}\ }\textbf {\bibinfo {volume} {19}},\ \bibinfo {pages} {2353} (\bibinfo
  {year} {2023})}\BibitemShut {NoStop}%
\bibitem [{\citenamefont {Dral}\ \emph {et~al.}(2017)\citenamefont {Dral},
  \citenamefont {Owens}, \citenamefont {Yurchenko},\ and\ \citenamefont
  {Thiel}}]{Dral2017}%
  \BibitemOpen
  \bibfield  {author} {\bibinfo {author} {\bibfnamefont {P.~O.}\ \bibnamefont
  {Dral}}, \bibinfo {author} {\bibfnamefont {A.}~\bibnamefont {Owens}},
  \bibinfo {author} {\bibfnamefont {S.~N.}\ \bibnamefont {Yurchenko}}, \ and\
  \bibinfo {author} {\bibfnamefont {W.}~\bibnamefont {Thiel}},\ }\href
  {\doibase 10.1063/1.4989536} {\bibfield  {journal} {\bibinfo  {journal} {The
  Journal of Chemical Physics}\ }\textbf {\bibinfo {volume} {146}},\ \bibinfo
  {pages} {244108} (\bibinfo {year} {2017})}\BibitemShut {NoStop}%
\bibitem [{\citenamefont {Hou}\ \emph {et~al.}(2023)\citenamefont {Hou},
  \citenamefont {Ge},\ and\ \citenamefont {Dral}}]{Hou2023}%
  \BibitemOpen
  \bibfield  {author} {\bibinfo {author} {\bibfnamefont {Y.-F.}\ \bibnamefont
  {Hou}}, \bibinfo {author} {\bibfnamefont {F.}~\bibnamefont {Ge}}, \ and\
  \bibinfo {author} {\bibfnamefont {P.~O.}\ \bibnamefont {Dral}},\ }\href
  {\doibase 10.1021/acs.jctc.2c01038} {\bibfield  {journal} {\bibinfo
  {journal} {Journal of Chemical Theory and Computation}\ } (\bibinfo {year}
  {2023}),\ 10.1021/acs.jctc.2c01038}\BibitemShut {NoStop}%
\bibitem [{\citenamefont {{Pinheiro Jr}}\ \emph {et~al.}(2021)\citenamefont
  {{Pinheiro Jr}}, \citenamefont {Ge}, \citenamefont {Ferr{\'{e}}},
  \citenamefont {Dral},\ and\ \citenamefont {Barbatti}}]{Pinheiro2021}%
  \BibitemOpen
  \bibfield  {author} {\bibinfo {author} {\bibfnamefont {M.}~\bibnamefont
  {{Pinheiro Jr}}}, \bibinfo {author} {\bibfnamefont {F.}~\bibnamefont {Ge}},
  \bibinfo {author} {\bibfnamefont {N.}~\bibnamefont {Ferr{\'{e}}}}, \bibinfo
  {author} {\bibfnamefont {P.~O.}\ \bibnamefont {Dral}}, \ and\ \bibinfo
  {author} {\bibfnamefont {M.}~\bibnamefont {Barbatti}},\ }\href {\doibase
  10.1039/d1sc03564a} {\bibfield  {journal} {\bibinfo  {journal} {Chemical
  Science}\ }\textbf {\bibinfo {volume} {12}},\ \bibinfo {pages} {14396}
  (\bibinfo {year} {2021})}\BibitemShut {NoStop}%
\bibitem [{\citenamefont {Barbatti}\ \emph {et~al.}(2022)\citenamefont
  {Barbatti}, \citenamefont {Bondanza}, \citenamefont {Crespo-Otero},
  \citenamefont {Demoulin}, \citenamefont {Dral}, \citenamefont {Granucci},
  \citenamefont {Kossoski}, \citenamefont {Lischka}, \citenamefont {Mennucci},
  \citenamefont {Mukherjee}, \citenamefont {Pederzoli}, \citenamefont
  {Persico}, \citenamefont {Jr}, \citenamefont {Pittner}, \citenamefont
  {Plasser}, \citenamefont {Gil},\ and\ \citenamefont
  {Stojanovic}}]{NewtonX2022}%
  \BibitemOpen
  \bibfield  {author} {\bibinfo {author} {\bibfnamefont {M.}~\bibnamefont
  {Barbatti}}, \bibinfo {author} {\bibfnamefont {M.}~\bibnamefont {Bondanza}},
  \bibinfo {author} {\bibfnamefont {R.}~\bibnamefont {Crespo-Otero}}, \bibinfo
  {author} {\bibfnamefont {B.}~\bibnamefont {Demoulin}}, \bibinfo {author}
  {\bibfnamefont {P.~O.}\ \bibnamefont {Dral}}, \bibinfo {author}
  {\bibfnamefont {G.}~\bibnamefont {Granucci}}, \bibinfo {author}
  {\bibfnamefont {F.}~\bibnamefont {Kossoski}}, \bibinfo {author}
  {\bibfnamefont {H.}~\bibnamefont {Lischka}}, \bibinfo {author} {\bibfnamefont
  {B.}~\bibnamefont {Mennucci}}, \bibinfo {author} {\bibfnamefont
  {S.}~\bibnamefont {Mukherjee}}, \bibinfo {author} {\bibfnamefont
  {M.}~\bibnamefont {Pederzoli}}, \bibinfo {author} {\bibfnamefont
  {M.}~\bibnamefont {Persico}}, \bibinfo {author} {\bibfnamefont {M.~P.}\
  \bibnamefont {Jr}}, \bibinfo {author} {\bibfnamefont {J.}~\bibnamefont
  {Pittner}}, \bibinfo {author} {\bibfnamefont {F.}~\bibnamefont {Plasser}},
  \bibinfo {author} {\bibfnamefont {E.~S.}\ \bibnamefont {Gil}}, \ and\
  \bibinfo {author} {\bibfnamefont {L.}~\bibnamefont {Stojanovic}},\ }\href
  {\doibase 10.1021/acs.jctc.2c00804} {\bibfield  {journal} {\bibinfo
  {journal} {Journal of Chemical Theory and Computation}\ }\textbf {\bibinfo
  {volume} {18}},\ \bibinfo {pages} {6851} (\bibinfo {year}
  {2022})}\BibitemShut {NoStop}%
\bibitem [{\citenamefont {Barbatti}\ \emph {et~al.}(2013)\citenamefont
  {Barbatti}, \citenamefont {Granucci}, \citenamefont {Ruckenbauer},
  \citenamefont {Plasser}, \citenamefont {Crespo-Otero}, \citenamefont
  {Pittner}, \citenamefont {Persico},\ and\ \citenamefont
  {Lischka}}]{NewtonX2013}%
  \BibitemOpen
  \bibfield  {author} {\bibinfo {author} {\bibfnamefont {M.}~\bibnamefont
  {Barbatti}}, \bibinfo {author} {\bibfnamefont {G.}~\bibnamefont {Granucci}},
  \bibinfo {author} {\bibfnamefont {M.}~\bibnamefont {Ruckenbauer}}, \bibinfo
  {author} {\bibfnamefont {F.}~\bibnamefont {Plasser}}, \bibinfo {author}
  {\bibfnamefont {R.}~\bibnamefont {Crespo-Otero}}, \bibinfo {author}
  {\bibfnamefont {J.}~\bibnamefont {Pittner}}, \bibinfo {author} {\bibfnamefont
  {M.}~\bibnamefont {Persico}}, \ and\ \bibinfo {author} {\bibfnamefont
  {H.}~\bibnamefont {Lischka}},\ }\href@noop {} {\enquote {\bibinfo {title}
  {{Newton-X} - package for newtonian dynamics close to the crossing seam, {\tt
  http://www.univie.ac.at/newtonx}},}\ } (\bibinfo {year}
  {2008-2013})\BibitemShut {NoStop}%
\bibitem [{TUR(2018)}]{TURBOMOLE_7.3}%
  \BibitemOpen
  \href@noop {} {\enquote {\bibinfo {title} {{TURBOMOLE V7.3 2018}, a
  development of {University of Karlsruhe} and {Forschungszentrum Karlsruhe
  GmbH}, 1989-2007, {TURBOMOLE GmbH}, since 2007; available from \\ {\tt
  http://www.turbomole.com}.}}\ } (\bibinfo {year} {2018})\BibitemShut
  {NoStop}%
\bibitem [{\citenamefont {Dral}(2019)}]{MLatom2019}%
  \BibitemOpen
  \bibfield  {author} {\bibinfo {author} {\bibfnamefont {P.~O.}\ \bibnamefont
  {Dral}},\ }\href {\doibase 10.1002/jcc.26004} {\bibfield  {journal} {\bibinfo
   {journal} {Journal of Computational Chemistry}\ }\textbf {\bibinfo {volume}
  {40}},\ \bibinfo {pages} {2339} (\bibinfo {year} {2019})}\BibitemShut
  {NoStop}%
\bibitem [{\citenamefont {Dral}\ \emph {et~al.}(2021)\citenamefont {Dral},
  \citenamefont {Ge}, \citenamefont {Xue}, \citenamefont {Hou}, \citenamefont
  {Pinheiro~Jr}, \citenamefont {Huang},\ and\ \citenamefont
  {Barbatti}}]{MLatom2021}%
  \BibitemOpen
  \bibfield  {author} {\bibinfo {author} {\bibfnamefont {P.~O.}\ \bibnamefont
  {Dral}}, \bibinfo {author} {\bibfnamefont {F.}~\bibnamefont {Ge}}, \bibinfo
  {author} {\bibfnamefont {B.-X.}\ \bibnamefont {Xue}}, \bibinfo {author}
  {\bibfnamefont {Y.-F.}\ \bibnamefont {Hou}}, \bibinfo {author} {\bibfnamefont
  {M.}~\bibnamefont {Pinheiro~Jr}}, \bibinfo {author} {\bibfnamefont
  {J.}~\bibnamefont {Huang}}, \ and\ \bibinfo {author} {\bibfnamefont
  {M.}~\bibnamefont {Barbatti}},\ }\href {\doibase 10.1007/s41061-021-00339-5}
  {\bibfield  {journal} {\bibinfo  {journal} {Topics in Current Chemistry}\
  }\textbf {\bibinfo {volume} {379}},\ \bibinfo {pages} {27} (\bibinfo {year}
  {2021})}\BibitemShut {NoStop}%
\end{thebibliography}%
